\documentclass[%
 reprint,
 amsmath,amssymb,
 aps,
showkeys
]{revtex4-2}

\usepackage{graphicx}
\usepackage{dcolumn}
\usepackage{bm}
\usepackage[colorlinks=true, linkcolor=blue, citecolor=blue, urlcolor=blue]{hyperref}
\usepackage{float} 

\begin{document}

\preprint{APS/123-QED}

\title{{\bf Eigenstate solutions of the Fermi-Hubbard model via symmetry-enhanced variational quantum eigensolver}  }

\author{Shaohui Yao\textsuperscript{1}}
\email{yaosh77@qq.com}
\author{Wenyu Wang\textsuperscript{1}}%
\affiliation{
 \textsuperscript{1}School of Physics and Optoelectronic Engineering, Beijing University of Technology, Beijing, China}

\date{\today}

\begin{abstract}
The Variational Quantum Eigensolver (VQE), as a hybrid 
quantum-classical algorithm, is an important tool for effective
quantum computing in the current noisy intermediate-scale
quantum (NISQ) era. However, the traditional hardware-efficient 
ansatz without taking into account symmetries requires 
more computational resources to explore the unnecessary regions 
in the Hilbert space. The conventional Subspace-Search VQE 
(SSVQE) algorithm, which can
calculate excited states, is also unable to effectively handle
degenerate states since the loss function only contains the 
expectation value of the Hamiltonian. In this study, the energy eigenstates of the one-dimensional Fermi-Hubbard model with two and four lattice sites are calculated, respectively. 
By incorporating symmetries into the quantum circuits 
and loss function, we find that both the
ground state and excited state calculations are improved greatly 
compared to the case without symmetries. 
The enhancement in excited state calculations is particularly
significant. This is because quantum circuits that conserve
the particle number are used, 
and appropriate penalty terms are added to the loss
function, enabling the optimization process to correctly identify
degenerate states. The results are verified
through repeated simulations.
\end{abstract}

\keywords{Quantum simulation, Hubbard model, Variational quantum eigensolver, Quantum computation}
\maketitle


\section{\label{sec:INTRODUCTION}INTRODUCTION}

The concept of quantum computing was first proposed by Feynman 
\cite{Feynman1982} due to the difficulties in the large scale 
quantum system simulation conducted on the classical computers.
Subsequently, Shor's algorithm, proposed by Shor based on quantum computers\cite{Shor1994}, attracted widespread attention to quantum computing for the first time, as it could break the Rivest-Shamir-Adleman (RSA) cryptosystem in polynomial time. This was also the first significant application of quantum computing beyond quantum simulation. Since quantum algorithms were first truly executed on a two-qubit experimental device \cite{Chuang1998} and Google company 
claimed to achieve quantum supremacy \cite{Google2019}, 
the impact of quantum computing has been growing up significantly.
Currently, limited by technology, experimental devices equipped with
qubits are referred to as noisy intermediate scale quantum (NISQ)
information processors 
\cite{NISQ2018, NISQMaterials2021, NISQChemistry2019}. 
This is due to the fact that quantum computers cannot scale to a
sufficient size. In other words,  quantum computers
cannot host a large enough number of qubits. 
Additionally, the presence of noise implies that 
the error effects must be taken into account.

On current NISQ devices, in order to harness the advantages of quantum
computing, the Variational Quantum Eigensolver (VQE) is proposed 
as a hybrid quantum-classical algorithm that
is capable of effectively utilizing quantum computing 
\cite{VQEOriginal2014, VQEOriginal2016, VQEReview2022, 
VQESimulation2023, VQEHubbard2022, VQEDMETHubbard2022, 
VQEMolecular2016, VQEMapping2023}.
It combines the advantages of quantum and classical computing,
respectively. VQE enables the implementation of quantum algorithms 
that have a clear advantage over classical algorithms when running
on a quantum computer. The optimization processes, currently 
beyond the capabilities of quantum computers, are handled by classical
computers. This collaboration between quantum and classical systems 
significantly enhances computational power.
The loss function in VQE algorithm for
the optimization is defined as the expectation
value of the system's Hamiltonian. The ground state of the system 
can be obtained by minimizing the loss function. 
Based on this algorithm, Nakanishi et al. further developed the
Subspace-Search VQE (SSVQE) algorithm \cite{SSVQE2019}, of which
the loss function is defined as a weighted sum of the Hamiltonian
expectation values over $n$ orthogonal initial states. 
Then the first $n$ energy eigenstates can be calculated
in case of non-degenerate states. 

For the quantum many-body systems, 
the dimension of the Hilbert space that describes the state
increases exponentially along with the increase of the number of 
particles in the system. The dimension of the Hamiltonian matrix
to be solved by classical computers also increases exponentially 
accordingly, making the computation blows up. 
This is also known as the ``exponential wall" problem 
\cite{ExponentialWall2010}. 
However, no matter VQE or SSVQE algorithm, 
the symmetries of the system, under which the unnecessary regions in
Hilbert space can be skipped during the optimization process, are not 
taken into account. Nevertheless, they fail to effectively distinguish
degenerate state in the calculation of excited states. This is
because of that the SSVQE algorithm only includes terms related to
the Hamiltonian expectation value. To resolve these issues, 
a symmetry-preserving method was developed by 
Lyu et. al \cite{SymmetryEnhancedVQE2023} and verified on the 
experimental devices \cite{SymmetryEnhancedVQE2024}. 
The hardware and hybrid symmetry-preserving methods were employed to
calculate the energy eigenstates of the one-dimensional Heisenberg model.
Note that symmetry is considered only in the quantum circuit in the
first one, while symmetry is included in both the quantum circuit 
and the loss function in the second one.

The Hubbard model is a model used in condensed matter physics to 
describe strongly correlated electron systems. Although the exact
solution to the one-dimensional Hubbard model has long been found
\cite{1DHubbard1992}, the solutions for the two-dimensional or 
three-dimensional Hubbard models have not yet been found. As an
important model for understanding the high-temperature 
superconducting mechanisms of materials like uprate~
\cite{CuprateHubbard2024}, 
solving the Hubbard model is of great significance.
As for the quantum many-body system, the Fermi-Hubbard model 
\cite{VQEHubbard2022, VQEDMETHubbard2022, HubbardModel2022, 
PanJianweiHubbard2024, LowDepthHubbard2019, TetrahedronHubbard1995,
MonteCarloHubbard1997} also faces the exponential wall problem 
along with the increase of the number of lattice sites.
The Hubbard Hamiltonian can transfer from fermionic creation and 
annihilation operators into a finite-dimensional matrix form by 
Jordan-Wigner transformation \cite{JordanWigner1928, JordanWigner2011}. 
The energy eigenstates of the system can be obtained by 
diagonalizing the matrix. 
However, the matrix dimension increases as $4^{n}$ with the number 
of lattice sites $n$. e.g. a matrix of size $4^{15} \approx 10^{9}$
needs to be solved in case of a system with 15 lattice sites.
This numerical calculation task far exceeds the capability of the most
powerful supercomputers currently in the world which can only handle 
eigenvalue problems up to $\approx 10^{8}$ dimensions 
\cite{Supercomputer2023}. 
The details of the evaluation are shown in Appendix 
~\ref{app:An_estimation}. 
In contrast, it is possible that just 30 qubits
are needed to solve the problem with the quantum computers. 

To reduce the scale of the calculation, symmetry 
can be taken into account in VQE algorithm for the Hubbard model.
In this study, regarding the quantum circuit, the invariance of the number of qubit excitations proposed in \cite{ParticleCircuit2020, ParticleCircuit2023,ParticleCircuit2004} is connected to the particle number symmetry of the Hubbard system and applied to realistic and physically meaningful systems. Moreover, symmetry is considered not only within the quantum circuit but also in the loss function, and is further utilized to solve degenerate excited states. Previous studies on solving the Hubbard model mostly focused on ground state solutions and did not emphasize the system’s symmetry.
In all, the paper is organized as following:
the VQE algorithm is introduced in Sec.~\ref{sec:VQE_Algorithm}.  
The Jordan-Wigner transformation is employed in Sec. 
~\ref{sec:Encoding_Operators}
to convert fermionic creation and annihilation operators into 
strings of Pauli matrices and to encode the Hubbard operators. 
The energy eigenstates of the system using the 
symmetry-enhanced VQE  algorithm
are studied in Sec.~\ref{sec:complete_eigenstate} and 
sec.~\ref{sec:partial_eigenstate}. 
Finally, our conclusions is given in sec.~\ref{sec:Conclusion}.

\vspace*{\fill}   

\section{\label{sec:VQE_Algorithm}Variational Quantum Eigensolver Algorithm}

\subsection{\label{sec:VQE_ground}VQE for ground state}
As talked in the introduction, 
due to hardware limitations in NISQ, quantum computers cannot
efficiently implement optimization algorithms.
However, the exponential growth of the Hilbert space dimension in 
quantum systems does not lead to computational difficulties for the
optimization process. Therefore, this step of the computation can be
performed on classical computers. 
The VQE algorithm precisely adopts this idea.

The algorithm can be roughly divided into two steps. 
In the first step, the state of the qubits is initialized
by a quantum circuit $U_{I}$ based on specific
requirements, yielding $\vert\psi_{I}\rangle=U_{I}
\left(\otimes_{i=0}^{n-1}\vert0\rangle_{i}\right)$, where $n$ is the number of
qubits. Then, the qubits are operated on by a parameterized quantum circuit 
$U\left(\boldsymbol{\theta}\right)$ with variable parameters, 
resulting in the state $\vert\psi\left(\boldsymbol{\theta}\right)\rangle=
U\left(\boldsymbol{\theta}\right)\vert\psi_{I}\rangle$. 
Measurements are performed to obtain this state. 
In the second step, the expectation value of the Hamiltonian with respect to the state $\vert\psi\left(\boldsymbol{\theta}\right)\rangle$ is calculated, given by $\langle\psi\left(\boldsymbol{\theta}\right)\vert \hat{H}\vert\psi\left(\boldsymbol{\theta}\right)\rangle$.
The loss function is defined as this expectation value,
\begin{equation}
\mathcal{L}\left(\boldsymbol{\theta}\right)=\langle\psi\left(\boldsymbol
{\theta}\right)\vert\hat{H}\vert\psi\left(\boldsymbol{\theta}\right)
\rangle\,.
\label{eq:loss_ground_state}
\end{equation}
An optimization algorithm, such as gradient descent 
\cite{GradientDescent2019}, is used to update the parameters.
The updated parameters are then passed to the quantum circuit, 
and such process is repeated. After multiple iterations, 
the minimized loss function is obtained, 
$\mathcal{L}(\boldsymbol{\theta}^{*})=\underset{\boldsymbol{\theta}}
{\text{min }}\mathcal{L}\left(\boldsymbol{\theta}\right)$, 
corresponding to the ground-state energy. 
The ground-state vector is obtained by applying the optimized 
quantum circuit to the initial state 
$\vert E_{0}\rangle=U(\boldsymbol{\theta}^{*})\vert\psi_{I}\rangle$.

For a single qubit, the variational parameters in the quantum circuit
represent the rotation angles of the qubit's state vector around the 
$x$, $y$, or $z$ axes on the Bloch sphere \cite{BlochSphere2019}. 
As for a two-state system 
$\vert\psi\rangle=\alpha\vert0\rangle+\beta\vert1\rangle$, 
it can be transformed in the two-dimensional complex vector space
by an  element $U\left(a,b\right)$ from the SU(2)
group, where $\left|a\right|^{2}+\left|b\right|^{2}=1$,
resulting in $\vert\psi^{'}\rangle=U\vert\psi\rangle$.
This corresponds to a rotational transformation in three-dimensional
real vector space. Since  SU(2) group and the SO(3) group are 
homomorphic, We can denote  that SO(3) group are represented 
by $a$, $b$ as $R\left(a,b\right)$ (the specific matrix representation is given in Appendix ~\ref{app:SO3_SU2}). Set 
$a=\mathrm{e}^{-\mathrm{i}\frac{\theta}{2}}$, $b=0$, then we have
\begin{equation}
R\left(a=\mathrm{e}^{-\mathrm{i}\frac{\theta}{2}},b=0\right)=
\begin{pmatrix}\cos\theta & -\sin\theta & 0\\
\sin\theta & \cos\theta & 0\\
0 & 0 & 1
\end{pmatrix}\,.
\end{equation}
This SO(3) element corresponds to a rotation around the $z$-axis. 
If $a=\mathrm{e}^{-\mathrm{i}\frac{\theta}{2}},b=0$ is substituted 
into the group element of the SU(2) group, we can get
\begin{eqnarray}
U\left(a^{'},b^{'}\right) & = & \begin{pmatrix}a & b\\
-b^{*} & a^{*}
\end{pmatrix}\Bigg|_{a=\mathrm{e}^{-\mathrm{i}\frac{\theta}{2}},b=0} \nonumber \\
& = & \begin{pmatrix}\mathrm{e}^{-\mathrm{i}\frac{\theta}{2}} & 0\\
0 & \mathrm{e}^{\mathrm{i}\frac{\theta}{2}}
\end{pmatrix}\,,
\end{eqnarray}
which corresponds to 
$R_{z}\left(\theta\right)=\mathrm{e}^{-\mathrm{i}\frac{\theta}{2}\sigma_{z}}$, a commonly used quantum gate. 
The variation of the group parameter $\theta$ corresponds to the
rotation of the vector around the $z$-axis in three-dimensional space.
Similarly, 
$R_{x}\left(\theta\right)=\mathrm{e}^{-\mathrm{i}\frac{\theta}{2}\sigma_{x}}$ 
and $R_{y}\left(\theta\right)=\mathrm{e}^{-\mathrm{i}\frac{\theta}{2}\sigma_{y}}$. 
This also implies that Pauli matrices $\sigma_{x}$, $\sigma_{y}$
and $\sigma_{z}$ are the generators of the state vector rotations
around the $x$-axis, $y$-axis and $z$-axis, respectively.

Note that the above algorithm is only for solving the ground state. The calculation for the excited states will be discussed in the next subsection.

\subsection{\label{sec:VQE_excited_state}VQE for excited state}

Although VQE is an effective algorithm, it is limited in calculating the
ground state and cannot solve for excited states. Nakanishi 
et al. \cite{SSVQE2019} developed a method, known as the Weighted 
SSVQE algorithm or simply the SSVQE algorithm, which can efficiently 
compute the first $k$ eigenstates of a system. This algorithm selects
a series of mutually orthogonal initial states and optimizes a loss
function defined as the weighted sum of the Hamiltonian's expectation
values for different initial states and weights, 
thereby obtaining the respective energy eigenstates.

Considering the case of non-degenerate states, $k$ mutually orthogonal states in the Hilbert space are initially constructed. 
In principle, rotating these states can make these states
approach $k$ energy eigenstates. 
Suppose these mutually orthogonal initial states are
$ \vert\phi_{0}\rangle, \vert\phi_{1}\rangle, \dots, \vert\phi_{k-1}\rangle$
 satisfying $\langle\phi_{i}\vert\phi_{j}\rangle=\delta_{ij}$,
 where $\delta_{ij}$ is the Kronecker delta function. 
 Define the corresponding loss function as
\begin{equation}
\mathcal{L}\left(\boldsymbol{\theta}\right)=\sum_{i=0}^{k-1}\omega_{i}\langle\phi_{i}\vert U^{\dagger}\left(\boldsymbol{\theta}\right)\hat{H}U\left(\boldsymbol{\theta}\right)\vert\phi_{i}\rangle\,,
\label{eq:loss_excited_state}
\end{equation}
where $\omega_{0}>\omega_{1}>\cdots>\omega_{k-1}$. 
By optimizing the loss function to approach the minimum value
of 
$\mathcal{L}(\boldsymbol{\theta}^{*})=\underset{\boldsymbol{\theta}}{\text{min }}\mathcal{L}\left(\boldsymbol{\theta}\right)$, 
we can obtain the energy eigenvalues 
\begin{eqnarray}
 E_{j}=\langle\phi_{j}\vert U^{\dagger}\left(\boldsymbol{\theta}^{*}\right)\hat{H}U\left(\boldsymbol{\theta}^{*}\right)\vert\phi_{j}\rangle,   
\end{eqnarray}
 and the corresponding energy eigenstates 
 \begin{eqnarray}
  \vert E_{j}\rangle=U\left(\boldsymbol{\theta}^{*}\right)\vert\phi_{j}\rangle \,,  
 \end{eqnarray}
  where $j\in\left\{0,1,\dots,k-1\right\}$. 
  To implement the VQE algorithm, it is necessary to encode the operators, which are described in the following section.

\section{\label{sec:Encoding_Operators}
Encoding of the Fermi-Hubbard Operators}

\subsection{\label{sec:Jordan-Wigner}The Jordan-Wigner transformation}

The Jordan-Wigner transformation was initially developed
to map the spin operators to fermionic creation and annihilation
operators \cite{JordanWigner1928}. Later, for the purposes of 
computation and measurement, it became necessary to transform
Hamiltonians expressed in terms of creation and annihilation operators
into combinations of Pauli operators \cite{JordanWigner2011}
in quantum computing. The process is exactly the inverse of the original transformation.
The raising and lowering operators for qubit states are 
defined as 
$\sigma^{-}=\vert0\rangle\langle1\vert=\frac{1}{2}\left(\sigma_{x}+\mathrm{i}\sigma_{y}\right)$ and $\sigma^{+}=\vert1\rangle\langle0\vert=\frac{1}{2}\left(\sigma_{x}-\mathrm{i}\sigma_{y}\right)$, where $\sigma_{x}$ and $\sigma_{y}$ are Pauli matrices. In the following, 
the identity matrix and Pauli matrices are denoted as $I$, $X$, $Y$ and $Z$.

The operator $\hat{a}_{i}$ represents the annihilation of a fermion
at the $i$-th lattice site. Suppose there are $N$ lattice sites, 
and each site's fermionic occupied and unoccupied states are 
mapped to the $\vert1\rangle$ and $\vert0\rangle$ states of a qubit, 
respectively. 
Intuitively, one may think that $\hat{a}_{i}$ can be transformed into  
\begin{equation}
\hat{a}_{i} = I_{0} \otimes I_{1} \otimes \cdots \otimes I_{i-1} \otimes \sigma_{i}^{-} \otimes I_{i+1} \otimes \cdots \otimes I_{N}\,,
\end{equation}
which applies the lowering operator $\sigma_{i}^{-}$ to the 
$i$-th qubit, while leaving all other qubits unchanged.  
However, this definition of the annihilation operator does not 
satisfy the fermionic anti-commutation relations: 
\begin{eqnarray}
 \left\{ \hat{a}_{k},\hat{a}_{l}^{\dagger}\right\} =\delta_{kl},\ \left\{ \hat{a}_{k},\hat{a}_{l}\right\} =0,\ \left\{ \hat{a}_{k}^{\dagger},\hat{a}_{l}^{\dagger}\right\} =0 \,.
\end{eqnarray}
The problem can be solved using the Jordan-Wigner transformation, which replaces all identity operators preceding the lowering operator for the $i$-th qubit with Pauli $Z$ operators, 
resulting in
\begin{equation}
\hat{a}_{i} = Z_{0} \otimes Z_{1} \otimes \cdots \otimes Z_{i-1} \otimes \sigma_{i}^{-} \otimes I_{i+1} \otimes \cdots \otimes I_{N}\,.
\end{equation}
The above is just the encoding for $\hat{a}_{i}$. Below, the method are applied to encode the Hubbard operators.

\subsection{\label{sec:Operators_Encoding}Operators Encoding}

The Fermi-Hubbard model considers only electron hopping between 
nearest-neighbor lattice sites and the Coulomb interaction between 
two electrons of opposite spin on the same site. 
Its Hamiltonian is given by
\begin{equation}
\hat{H} = -t\sum_{i,j}\sum_{\sigma}\left(\hat{a}_{i\sigma}^{\dagger}\hat{a}_{j\sigma} + \hat{a}_{j\sigma}^{\dagger}\hat{a}_{i\sigma}\right) + U\sum_{i}\hat{n}_{i\uparrow}\hat{n}_{i\downarrow}\,,
\label{eq:Hamiltonian}
\end{equation}
where $\sigma \in \{\uparrow, \downarrow\}$ 
and $\hat{n}_{i\sigma} = \hat{a}_{i\sigma}^{\dagger}\hat{a}_{i\sigma}$.
The term $\hat{a}_{i\sigma}^{\dagger}\hat{a}_{j\sigma}$ represents 
the annihilation of an electron with spin $\sigma$ at site $j$ and 
the creation of an electron with spin $\sigma$ at site $i$, 
describing electron hopping between sites $i$ and $j$. T
he summation $\sum_{i,j}\sum_{\sigma}$ accounts for all the 
nearest-neighbor sites and both spin states. Here, $t$ is the hopping
amplitude, indicating the probability of electron hopping between 
two lattice sites. The operator 
$\hat{n}_{i\sigma} = \hat{a}_{i\sigma}^{\dagger}\hat{a}_{i\sigma}$
is the particle number operator for electrons with spin $\sigma$. 
$\hat{n}_{i\uparrow}\hat{n}_{i\downarrow}$ denotes the number of electron pairs with opposite spins on site $i$ 
(taking values of 0 or 1). Finally, $U$ represents the Coulomb 
interaction potential between two electrons of opposite spin on the
same lattice site. In this study, we choose $t=1$ and $U=2$. 
This is a common choice \cite{HubbardtU2015, HubbardtU2020}.

Considering that each electron has two spin eigenstates, the occupation
and vacant states of spin-up electrons on $N$ lattice sites are mapped
to the two eigenstates of the first $ N $ qubits. Similarly, the 
occupation and vacant states of spin-down electrons on $ N $ lattice 
sites are mapped to the two eigenstates of the qubits from $ N+1 $ to 
$ 2N $. The encoding of $\hat{a}_{i\uparrow}$, the annihilation 
operator for spin-up fermions on the $ i $-th lattice site, 
in a lattice with $ N $ sites is
\begin{eqnarray}
\hat{a}_{i\uparrow} & = & Z_{0}\otimes Z_{1}\otimes\cdots\otimes Z_{i-1}\otimes\sigma_{i}^{-}\otimes I_{i+1}\otimes I_{i+2} \nonumber \\
 &  & \otimes\cdots\otimes I_{N-1}\otimes I_{N}\otimes I_{N+1}\otimes\cdots\otimes I_{2N-1} \nonumber \\
 & = & \prod_{k=0}^{i-1}Z_{k}\otimes\sigma_{i}^{-}\otimes\prod_{k=i+1}^{2N-1}I_{k}\,,
\end{eqnarray}
where $\prod_{k=m}^{n} O_{k}$ denotes the sequential tensor product 
of operators, i.e., 
$\prod_{k=m}^{n} O_{k} = O_{m} \otimes O_{m+1} \otimes \cdots 
\otimes O_{n}$. The subscripts in the operators indicate the qubits 
on which the operators act. The term $I_{N} \otimes I_{N+1} \otimes \cdots \otimes I_{2N-1}$ implies no operations are performed on 
qubits $ N+1 $ to $ 2N $. It means that no spin-down electrons 
are created or annihilated. Similarly, the encodings for 
$\hat{a}_{i\uparrow}^{\dagger}$, $\hat{a}_{i\downarrow}$, and
$\hat{a}_{i\downarrow}^{\dagger}$ can be derived 
(see Appendix ~\ref{app:Derivation_steps}).

Through the matrix multiplication, the Jordan-Wigner encoding of the Hubbard model Hamiltonian is given by
\begin{eqnarray}
\hat{a}_{i\uparrow}^{\dagger}\hat{a}_{j\uparrow}+\hat{a}_{j\uparrow}^{\dagger}\hat{a}_{i\uparrow} & = & \frac{1}{2}\otimes\left(X_{i}\otimes\prod_{k=i+1}^{j-1}Z_{k}\otimes X_{j}\right. \nonumber \\
 &  & \left.+Y_{i}\otimes\prod_{k=i+1}^{j-1}Z_{k}\otimes Y_{j}\right) \,,\\
\hat{a}_{i\downarrow}^{\dagger}\hat{a}_{j\downarrow}+\hat{a}_{j\downarrow}^{\dagger}\hat{a}_{i\downarrow} & = & \frac{1}{2}\otimes\left(X_{i+N}\otimes\prod_{k=i+1+N}^{j-1+N}Z_{k}\otimes X_{j+N}\right. \nonumber \\
 &  & \left.+Y_{i+N}\otimes\prod_{k=i+1+N}^{j-1+N}Z_{k}\otimes Y_{j+N}\right) \,,
\end{eqnarray}

\begin{eqnarray}
\hat{n}_{i\uparrow}\hat{n}_{i\downarrow} & = & \frac{1}{4}\otimes\left(I_{i}\otimes\prod_{k=i+1}^{i-1+N}I_{k}\otimes I_{i+N}\right. \nonumber \\
 &  & -I_{i}\otimes\prod_{k=i+1}^{i-1+N}I_{k}\otimes Z_{i+N} \nonumber \\
 &  & -Z_{i}\otimes\prod_{k=i+1}^{i-1+N}I_{k}\otimes I_{i+N} \nonumber \\
 &  & \left.+Z_{i}\otimes\prod_{k=i+1}^{i-1+N}I_{k}\otimes Z_{i+N}\right) \,.
\end{eqnarray}
For simplicity, some identity matrices $I$ are 
omitted here, but these identity matrices cannot be ignored when 
performing simulations. The complete expressions 
can be found in Appendix ~\ref{app:Derivation_steps}.

The Hubbard Hamiltonian commutes with the particle number operator
$\hat{N}$ and the $z$-component of the total spin operator
$\hat{S}_{z}$ in this system \cite{SymmetryHubbard2022, SymmetryHubbard1989}:
\begin{eqnarray}
\left[\hat{H},\hat{N}\right] & = & 0 \,,\\
\left[\hat{H},\hat{S}_{z}\right] & = & 0\,,
\end{eqnarray}
where $\hat{N} = \sum_{i,\sigma} \hat{a}_{i\sigma}^{\dagger}\hat{a}_{i\sigma}$ and $\hat{S}_{z} = \frac{1}{2} \sum_{i} \left(\hat{n}_{i\uparrow} - \hat{n}_{i\downarrow}\right)$. 
This implies that the particle number operator and the $z$-component 
of the total spin operator share common eigenstates with the
Hamiltonian. In this paper, these two symmetries
are combined in the VQE algorithm.

The encoding of these two operators can be derived from the 
encoding of $\hat{a}_{i\uparrow}$, $\hat{a}_{i\uparrow}^{\dagger}$, $\hat{a}_{i\downarrow}$, and $\hat{a}_{i\downarrow}^{\dagger}$
obtained above. The detailed calculations are shown in the Appendix
~\ref{app:Derivation_steps}. In the next section, all eigenstates of
the one-dimensional Fermi-Hubbard model with two lattice sites 
are calculated based on these encoding.

\section{\label{sec:complete_eigenstate}The complete eigenstate
solutions of the two-site 1D Hubbard model}

\subsection{\label{sec:Method_of_implementation}Method of implementation}

First, we introduce how quantum computing simulations are performed. 
In the VQE algorithm, the quantum circuit that operates on qubits 
should be implemented on a quantum computer. However, as for
a theoretical exploration in this study 
the simulation are performed on a classical computer 
instead of that on a real quantum computer. All quantum states,
operators, and quantum circuits are converted into matrix forms,
and quantities such as expectation values and fidelity are calculated
by using matrix-related operations.

The states $\vert0\rangle$ and $\vert1\rangle$ are represented
as matrices as follows
\begin{equation}
\vert0\rangle=\begin{pmatrix}1\\
0
\end{pmatrix},\quad \vert1\rangle=\begin{pmatrix}0\\
1
\end{pmatrix}\,.
\end{equation}
This is the standard representation in quantum computing research. 
The operation of the quantum gates in each column of a quantum circuit
on qubits can be expressed as 
$G_{c0}\left(\boldsymbol{\theta}\right)=G_{0}\otimes 
G_{1}\otimes\cdots\otimes G_{r}$, where $G_{i}$ is the matrix 
representing a single-qubit or multi-qubit gate.
These matrices satisfy $\dim\left(G_{i}\right)=2^{k}\times2^{k}$, 
where $k$ is the number of qubits the gate acts on. 
The quantum circuit represented in matrix form is given by
$U_{m}\left(\boldsymbol{\theta}\right)=G_{c0}G_{c1}\cdots G_{cn}$,
where $n$ is the number of columns of quantum gates 
in the circuit, and $G_{ci}$ are multiplied as matrices.

Next, we calculate the expectation value of an operator 
$\langle\psi\vert U_{m}^{\dagger}\left(\boldsymbol{\theta}\right)\hat{O}U_{m}\left(\boldsymbol{\theta}\right)\vert\psi\rangle$.
The intermediate term are calculated at first
$\hat{O}\left(\boldsymbol{\theta}\right)=U_{m}^{\dagger}\left(\boldsymbol{\theta}\right)\hat{O}U_{m}\left(\boldsymbol{\theta}\right)$.
After that, $\langle\psi\vert\hat{O}\left(\boldsymbol{\theta}\right)
\vert\psi\rangle$
are calculated. 
These calculations are matrix multiplications. 
On a real quantum computer, the state should  measured first
to obtain $\vert\psi\left(\boldsymbol{\theta}\right)\rangle$.
Then the expectation value of the operator $\hat{O}$ are computed 
out on this state, $\langle\psi\left(\boldsymbol{\theta}\right)\vert\hat{O}\vert\psi\left(\boldsymbol{\theta}\right)\rangle$. 
Note that the approach of calculating expectation values through 
matrix multiplication is equivalent to performing an ideal measurement, that is, a measurement with no errors and a sufficiently large number of repetitions, yielding the theoretical result.
In the following two subsections, we use this method to compute the
ground state and excited states, respectively.

\subsection{\label{sec:Eigenstate_solutions} Solutions 
of eigenstate via symmetry-enhanced VQE}

\begin{figure*}
\includegraphics[width=1.0\textwidth]{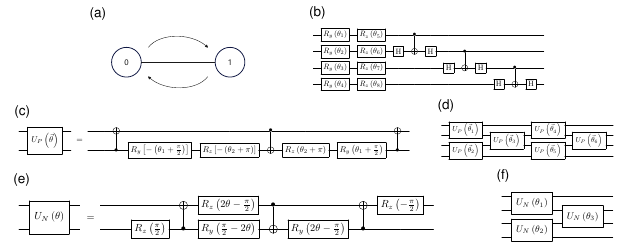}
\caption{\label{fig:lattice_and_quantum_circuits} (a) A one-dimensional
Hubbard model with two lattice sites, in which  the circles represent
the lattice sites, the numbers 0 and 1 in the center are the site
indices, and the arrows indicate the electron hopping between the
lattice sites. (b) Hardware efficient ansatz single-layer
circuits acting on four qubits. During the computation, $n$ layers of
such circuits will be applied. In a certain range, increasing the
number of layers can enhance the circuit's ability to represent the
state. (c) Quantum circuits that can conserve the particle
number for states such as $\vert01\rangle$ or $\vert10\rangle$. 
(d) The single-layer circuit corresponding to the quantum gate
in plot (c). (e) Quantum circuits that can conserve the particle number for arbitrary two-qubit state. (f) The single-layer circuit corresponding to the quantum gate in plot (e).}
\end{figure*}

\subsubsection{Ground state}

\begin{figure*}
\includegraphics[width=1.0\textwidth]{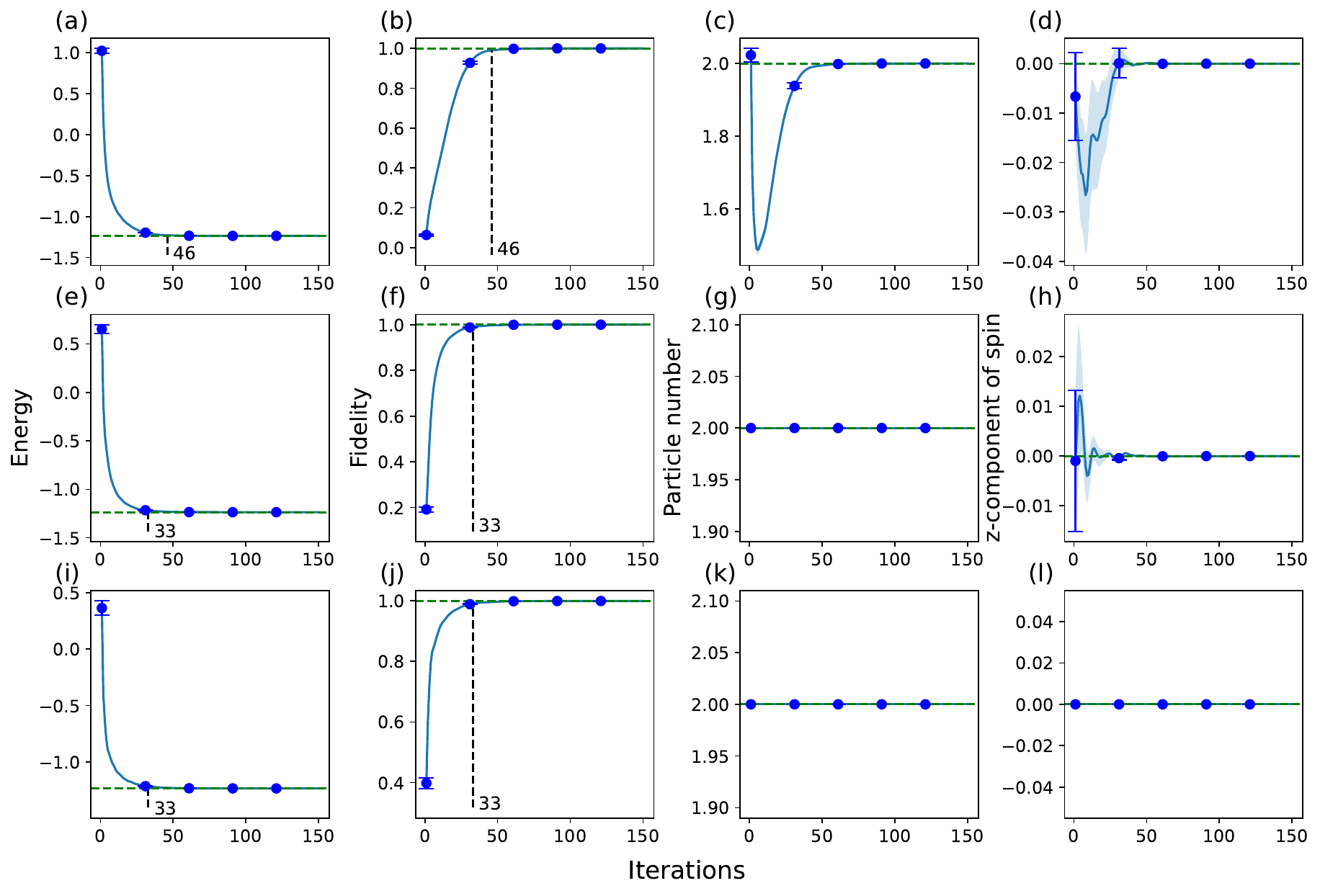}
\caption{\label{fig:1D_ground_state} 
The solved ground states under three different ansatzes. 
All data represent the average values from 200 independent 
computational runs, during which the initial parameters of the quantum 
circuit are randomly selected.  The error bands are displayed in blue
shade. The standard error is small and not visually prominent in
several plots. 
To make the standard error more apparent, error bars are provided at
every 30th iteration. Plots (a)-(d) show the energy expectation value
$\langle\psi_{0}\left(\boldsymbol{\theta}\right)\vert\hat{H}\vert\psi_{0}\left(\boldsymbol{\theta}\right)\rangle$, 
fidelity $\left|\langle\psi_{0}\left(\boldsymbol{\theta}\right)\vert\alpha_{2,0}\rangle\right|^{2}$, 
particle number expectation value $\langle\psi_{0}\left(\boldsymbol{\theta}\right)\vert\hat{N}\vert\psi_{0}\left(\boldsymbol{\theta}\right)\rangle$, 
and total spin operator $z$-component $\langle\psi_{0}\left(\boldsymbol{\theta}\right)\vert\hat{S}{z}\vert\psi_{0}\left(\boldsymbol{\theta}\right)\rangle$ 
as a function of iteration number using hardware efficient ansatz 
to solve for the ground state. Plot (c) shows that certain optimization
process is required to make the particle number approach the value 2,
which suggests that resources are being spent on searching for 
unnecessary states. Plots (e)-(h) show the results corresponding to 
the first type of symmetry-preserving ansatz. 
Plot (g) indicates that this ansatz is able to preserve the 
particle number. Plots (i)-(l) show the results obtained by using the
second type of symmetry-preserving ansatz. Plots (k)-(l) indicate 
that the circuit can simultaneously preserve both the particle number
and the $z$-component of spin when the initial state has a 
$z$-component of spin equal to zero. This eliminates the 
requirement 
to search for states that do not satisfy the target energy 
eigenstate, particle number, or $z$-component of spin.}
\end{figure*}

\begin{table}[b]
\caption{\label{tab:eigenstates}
Eigenstates with corresponding energy, particle number, and $z$-component of the total spin for a two-site Hubbard model.
}
\begin{ruledtabular}
\begin{tabular}{lccc}
\textrm{Eigenstate} & 
\textrm{Energy} & 
\textrm{Particle number} & 
\textrm{$z$-component of spin} \\
\colrule
$\vert\alpha_{2,0}\rangle$    & $-1.2361$ & $2$ & $0$ \\
$\vert\beta_{1,-\frac{1}{2}}\rangle$ & $-1.0000$ & $1$ & $-\frac{1}{2}$ \\
$\vert\beta_{1,\frac{1}{2}}\rangle$  & $-1.0000$ & $1$ & $\frac{1}{2}$ \\
$\vert\gamma_{2,-1}\rangle$   & $0.0000$  & $2$ & $-1$ \\
$\vert\gamma_{0,0}\rangle$    & $0.0000$  & $0$ & $0$ \\
$\vert\gamma_{2,1}\rangle$    & $0.0000$  & $2$ & $1$ \\
$\vert\gamma_{2,0}\rangle$    & $0.0000$  & $2$ & $0$ \\
$\vert\delta_{3,-\frac{1}{2}}\rangle$ & $1.0000$  & $3$ & $-\frac{1}{2}$ \\
$\vert\delta_{1,\frac{1}{2}}\rangle$  & $1.0000$  & $1$ & $\frac{1}{2}$ \\
$\vert\delta_{1,-\frac{1}{2}}\rangle$ & $1.0000$  & $1$ & $-\frac{1}{2}$ \\
$\vert\delta_{3,\frac{1}{2}}\rangle$  & $1.0000$  & $3$ & $\frac{1}{2}$ \\
$\vert\varepsilon_{2,0}\rangle$ & $2.0000$  & $2$ & $0$ \\
$\vert\zeta_{3,-\frac{1}{2}}\rangle$  & $3.0000$  & $3$ & $-\frac{1}{2}$ \\
$\vert\zeta_{3,\frac{1}{2}}\rangle$   & $3.0000$  & $3$ & $\frac{1}{2}$ \\
$\vert\eta_{2,0}\rangle$      & $3.2361$  & $2$ & $0$ \\
$\vert\theta_{4,0}\rangle$    & $4.0000$  & $4$ & $0$ \\
\end{tabular}
\end{ruledtabular}
\end{table}

In the simulation, 
the case of a lattice with two lattice sites was considered first. 
These sites are numbered as 1 and 2, 
as shown in Fig.~\ref{fig:lattice_and_quantum_circuits}. 
The arrows represent the electron transitions between the lattice
sites. The system has 16 energy eigenstates, 
denoted as $\vert\alpha\rangle$, $\vert\beta\rangle$,
$\vert\gamma\rangle$, etc., representing the ground state, the first
excited state, the second excited state, and so on. 
The subscripts indicate the eigenvalues of the particle number 
operator and $z$-component of the total spin operator for each
eigenstate. For instance, $\vert\alpha_{2,0}\rangle$ denotes
the ground state with eigenvalues $\alpha_{2,0}$. All the energy
eigenstates and the corresponding eigenvalues of the operators are
given in Table~\ref{tab:eigenstates}. These eigenstate information 
will be used in constructing the initial state and the loss function.

When constructing the initial state, it is necessary to know the
particle number and the $z$-component of spin corresponding to 
that state. Therefore, we need to determine the eigenvalue equations 
of these two operators in the computational basis. 
The following analysis provides the details.
The two eigenstates of the qubit correspond to the encoding of the 
spin-up and spin-down occupation states and vacant states of electrons 
at each lattice site. Therefore, it is evident that each computational 
basis corresponds to the eigenstates of the $S_{z}$ operator. 
Qubits 0-1 encode the occupation states and vacant states of spin-up 
electrons. The contribution to the $S_{z}$ eigenvalue is 0 when the
qubit state is $\vert0\rangle$ and $\frac{1}{2}$ when the qubit state
is $\vert1\rangle$. Qubits 2-3 encode the occupation states and vacant
states of spin-down electrons. The contribution to the $S_{z}$
eigenvalue is 0 when the qubit state is 
$\vert0\rangle$ and $-\frac{1}{2}$ when the qubit state is 
$\vert1\rangle$. 
For example,  the eigenvalue of $\hat{S}_{z}$ operator 
for the state $\vert0_{0}1_{1}1_{2}1_{3}\rangle$
is $\left(\frac{1}{2}\right)_{1}+\left(-\frac{1}{2}\right)_{2}
+\left(-\frac{1}{2}\right)_{3}=-\frac{1}{2}$, 
where the subscript indicates the qubit index.

Based on the above analysis, 
the eigenvalue equation for the $\hat{S}_{z}$ operator can be derived 
\begin{equation}
\hat{S}_{z}\vert\varphi_{z}\rangle=M\vert\varphi_{z}\rangle\,,
\end{equation}
where $M=\sum_{k}s_{k},\ i=0,1,\ j=2,3,\ k=i,j$. 
When the state of the $i$-th qubit is $\vert0\rangle$, $s_{i}=0$. 
Similarly, when $\vert q_{i}\rangle=\vert1\rangle$,
$s_{i}=\frac{1}{2}$; when $\vert q_{j}\rangle=\vert0\rangle$, $s_{j}=0$;
and when $\vert q_{j}\rangle=\vert1\rangle$, $s_{j}=-\frac{1}{2}$.
$\vert\varphi_{z}\rangle\in\left\{ 
\vert0000\rangle,\vert0001\rangle,...,\vert1111\rangle\right\} $. 
The same applies to the particle number operator $\hat{N}$.

To solve the ground state, 
information related to the ground state from 
above analysis is utilized. 
Using the VQE algorithm with the loss function defined in 
Eq.~(\ref{eq:loss_ground_state}),
the ground state of the system can be obtained. 
The hardware efficient ansatz without symmetry preservation and
two different symmetry-preserving ansatz are adopted 
for the comparison. The
computational resources required are measured by the number of
parameters in the quantum circuit, the number of CNOT gates, and the 
number of iterations in the optimization process etc. 
For all these three ansatz, we use the same initial state,  
\begin{equation}
\vert\psi_{0}\rangle=\vert\psi^{+}\rangle\otimes\vert\psi^{+}\rangle\,,
\end{equation}
where $\vert\psi^{+}\rangle=\frac{1}
{\sqrt{2}}\left(\vert01\rangle+\vert10\rangle\right)$. 
This initial state has the same particle number and 
$z$-component of spin as the ground state, enabling quantum circuits 
that preserve particle number to find the ground state more efficiently.

The quantum circuit for the hardware efficient ansatz is shown in
Fig.~\ref{fig:lattice_and_quantum_circuits}(b), in which 
one layer of the circuit is depicted. 12 layers of the circuit
are used to generate the ground state, which contains 96 parameters 
and 36 CNOT gates. The process are  repeated for 200 times. 
Each data point in the blue solid line corresponding to the iteration 
count in Fig.~\ref{fig:1D_ground_state} represents the average value 
of these 200 repetitions, and  the standard error and provided error
bands and error bars are also calculated.
In these 200 repetitions, the first iteration where
the average fidelity $\left|\langle\psi_{0}\left(\boldsymbol{\theta}\right)\vert\alpha_{2,0}\rangle\right|^{2}$ is greater than or equal to 0.99 
corresponds to iteration 46.

The first symmetry-preserving ansatz we use consists of two-qubit 
gates as the basic unit \cite{ParticleCircuit2020, ParticleCircuit2023},  
\begin{equation}
U_{P}\left(\boldsymbol{\theta}\right)=\begin{pmatrix}1 & 0 & 0 & 0\\
0 & \cos\theta_{1} & \mathrm{e}^{\mathrm{i}\theta_{2}}\sin\theta_{1} & 0\\
0 & \mathrm{e}^{-\mathrm{i}\theta_{2}}\sin\theta_{1} & -\cos\theta_{1} & 0\\
0 & 0 & 0 & 1
\end{pmatrix}\,.
\end{equation}
This two-qubit gate can preserve the particle number of states 
like $\vert01\rangle$ or $\vert10\rangle$. For example, 
$U_{P}\left(\boldsymbol{\theta}\right)\vert01\rangle=\cos\theta_{1}\vert01\rangle+\mathrm{e}^{-\mathrm{i}\theta_{2}}\sin\theta_{1}\vert10\rangle$. 
This is satisfied for the initial state we selected. 
However, this gate can only preserve the particle number for such
states and does not preserve it for states like $\vert00\rangle$ or 
$\vert11\rangle$. This limitation means that this quantum gate cannot
be used as a symmetry-preserving method to solve for other eigenstates. 
The quantum circuit corresponding to this gate is shown in 
Fig.~\ref{fig:lattice_and_quantum_circuits}(c). The quantum circuit 
for applying it to four qubits in a single layer is shown in 
Fig.~\ref{fig:lattice_and_quantum_circuits}(d). 
Two layers of this circuit are used to generate the ground state, 
containing 48 parameters and 36 CNOT gates. 
The first iteration where the average fidelity $\geq0.99$ occurs 
at iteration 33, showing a noticeable improvement over the 
hardware efficient ansatz. This is because the symmetry-preserving 
ansatz keeps the particle number fixed, so we do not 
have to search  the state space for states that do not match the
required particle number. Figure~\ref{fig:1D_ground_state}(g) shows 
the expectation value of the particle number operator 
$\langle\psi_{0}\left(\boldsymbol{\theta}\right)\vert\hat{N}\vert\psi_{0}\left(\boldsymbol{\theta}\right)\rangle$,
which remains constant throughout the computation. 
In contrast, for the hardware efficient ansatz, 
Fig.~\ref{fig:1D_ground_state}(c) requires optimization over a large
state space to reach the ground state with the correct particle number.

The other symmetry-preserving ansatz we use consists of two-qubit 
gates as the basic unit 
\cite{ParticleCircuit2004},  
\begin{equation}
U_{N}\left(\theta\right)=\mathrm{e}^{\mathrm{i}\theta\left(\sigma_{x}\otimes\sigma_{x}+\sigma_{y}\otimes\sigma_{y}+\sigma_{z}\otimes\sigma_{z}\right)}\,.
\end{equation}
This two-qubit gate can preserve the particle number of the initial
state regardless of its type. The quantum circuit corresponding to this 
gate and its single-layer version applied to four qubits are shown 
in Fig.~\ref{fig:lattice_and_quantum_circuits}(e) and
Fig.~\ref{fig:lattice_and_quantum_circuits}(f), respectively. 
Four layers of this circuit are used to generate the ground state, 
which includes 12 parameters and 36 CNOT gates. The first iteration 
where the average fidelity $\geq0.99$ occurs at iteration 33.
Similarly, the results show a noticeable improvement over the hardware
efficient ansatz. The circuit preserves the $z$-component of spin for
states with an initial $z$-component of spin equal to 0. 
This conclusion is evident from our simulation results.

In the comparison process, the number of CNOT gates is kept the same 
for the three different ansatz. However, the symmetry-preserving ansatz
require significantly fewer parameters than the hardware efficient 
ansatz. Specifically, the second symmetry-preserving ansatz requires
only 12 parameters, while the hardware efficient ansatz and the first 
symmetry-preserving ansatz require 96 and 48 parameters, respectively. 
Under these conditions, the number of iterations required to generate 
the ground state is still smaller for the symmetry-preserving ansatz
compared to the hardware efficient ansatz. This indicates that the 
symmetry-preserving methods reduce the computational resources required
and improve the efficiency of generating the ground state. 
Below, we will apply symmetry method to solve for the excited states.

\subsubsection{Excited states}
\begin{figure*}
\includegraphics[width=1.0\textwidth]{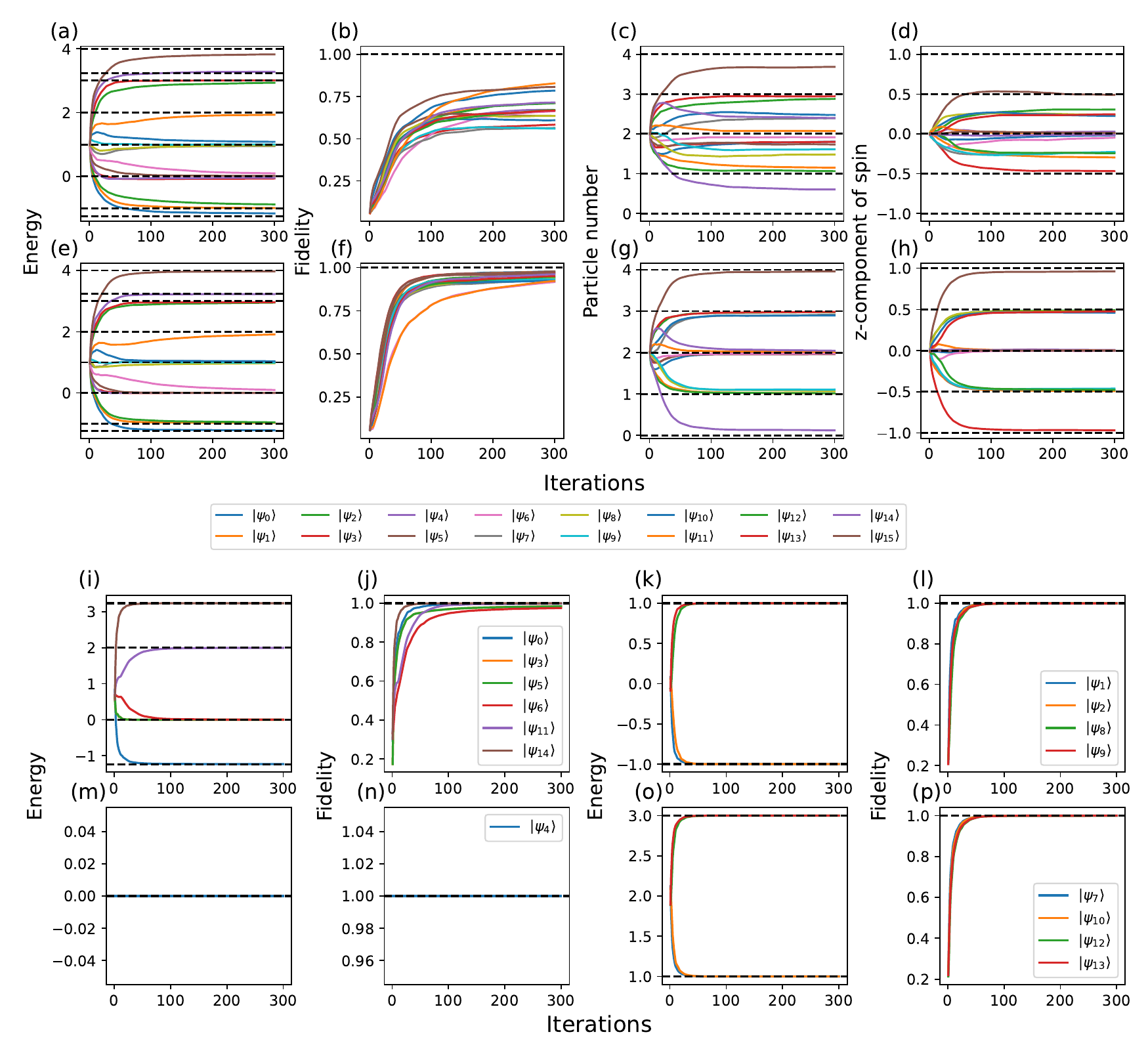}
\caption{\label{fig:1D_all_eigenstates} 
The solutions of all eigenstates obtained by using three different 
methods. The expectation values of various quantities during the 
computation process, with $\vert\psi_{i}\rangle$ as the initial state, 
are presented. The eigenstates corresponding to $\vert\psi_{i}\rangle$,
which need to be solved, are listed in 
Table~\ref{tab:eigenstates_initial_states}. All data represent the 
averages of 200 computations with random initial parameters. (a)-(d) 
show the changes in energy, fidelity, particle number, and $z$-component
of spin with respect to the number of iterations when using hardware 
efficient ansatz without adding penalty terms to the loss function. 
The fidelity plot in (b) demonstrates the inefficacy of this method in 
calculating degenerate states. Although the expectation value of the 
Hamiltonian in the loss function allows the identification of states 
with the correct energy, it fails to select one of the degenerate 
states. As a result, the energy in (a) converges well, but the particle 
number and $z$-component of spin in (c)-(d) fail to converge to the 
correct values. (e)-(h) show the changes of various quantities with
respect to the number of iterations when using the hardware efficient
ansatz with penalty terms. This method allows us to select which
degenerate state to be optimized, resulting in a high fidelity. 
(i)-(p) show the results using the symmetry-preserving ansatz with
penalty terms. This ansatz preserves the particle number, allowing 
faster convergence to the target energy eigenstate. 
(i)-(j), (k)-(l), (m)-(n) and (o)-(p) share the same legend, 
respectively. The fidelity in (n) starts at 1 because the initial 
state which we chose to satisfy the conditions is exactly the 
corresponding energy eigenstate. Similarly, the energy 
$\langle\psi_{15}\left(\boldsymbol{\theta}\right)\vert\hat{H}\vert\psi_{15}\left(\boldsymbol{\theta}\right)\rangle$ and the fidelity 
$\left|\langle\psi_{15}\left(\boldsymbol{\theta}\right)\vert\theta_{4,0}\rangle\right|^{2}$ corresponding to the state $\vert\psi_{15}\rangle$ remain consistently 4 and 1, respectively. 
For simplicity, they are not included in the figure.}
\end{figure*}

The SSVQE algorithm are used to generate all the excited states of this
system. The 16 selected mutually orthogonal initial states are listed 
in Appendix ~\ref{app:computational_resources}. First, we apply the 
hardware efficient ansatz using a 15-layer quantum circuit as shown in 
Fig.~\ref{fig:lattice_and_quantum_circuits}(b). 
For the conventional SSVQE, the loss function is defined as 
Eq.~(\ref{eq:loss_excited_state}). For non-degenerate states,
this loss function can effectively solve for the first $n$ 
eigenstates of the system. However, it is not suitable for solving
degenerate states. For several degenerate states with the same energy,
using the energy expectation value as a term in the loss function 
alone allows the eigenstate corresponding to that term 
to be any one of the degenerate states.

Figs.~\ref{fig:1D_all_eigenstates}(a)-(d) show the energy values, 
fidelities, particle numbers, and $z$-components of spin corresponding 
to each state solved during this process. The data for all the figures 
represent the averages over 200 optimization runs. When considering only the obtained energy values, they closely align with the theoretical results. However, examining the
fidelities of the corresponding states reveals that this method 
is not as effective. With this approach, we cannot choose which 
degenerate state to optimize. When taking the inner product with the 
theoretical state and calculating the square of the modulus, 
it may yield a result close to 0 because degenerate states are 
orthogonal to each other. Of course, it is also possible to obtain 
fidelities close to 1, leading to the averaged fidelity results shown 
in the figure over multiple runs. To converge to a specific eigenstate, 
the state must have the correct energy, particle number, and 
$z$-component of spin. While this method yields energy values that 
are relatively accurate, the particle number and $z$-component of 
spin significantly deviate from the theoretical values, 
as shown in Figs.~\ref{fig:1D_all_eigenstates}(c)-(d). 
For example, the eigenstate $\vert\gamma_{2,1}\rangle$ has a 
$z$-component of spin equal to 1, whereas the obtained state 
$\vert\psi_{5}(\boldsymbol{\theta}^{*})\rangle$ 
has a $z$-component of spin approximately equal to 0.5.

If a penalty term are added to the loss function, it enables the
algorithm to find states that minimize both the energy term and the penalty term simultaneously. The loss function is defined as  
\begin{equation}
\mathcal{L}=\sum_{i=0}^{k-1}\omega_{i}\left[\langle\hat{H}\rangle_{i}+\beta_{i}\left(\langle\hat{S}_{z}\rangle_{i}-M_{i}\right)^{2}\right]\,.
\label{eq:loss_hardware_Sz}
\end{equation}
The specific expression can be found in Appendix
~\ref{app:computational_resources}. The penalty factor $\beta$ must 
be chosen appropriately. If it is too small, the weight of the penalty 
term will be too low, and the algorithm will prioritize optimizing the
energy term, making it difficult to find states with the target spin. 
However, $\beta$ cannot be too large either, as an excessively high 
weight for the penalty term will dominate the optimization process,
slowing convergence to the target energy state. Here, we choose $\beta$ 
to be 0.5. When the $z$-component of spin is 0, the corresponding state 
can be found without adding a penalty term since other degenerate 
states with the same energy are penalized. Furthermore, in the 
symmetry-preserving quantum circuits used later, when the $z$-component 
of spin is 0, the circuit not only preserves the particle number but
also keeps the $z$-component of spin unchanged. 
Therefore, when solving for such states, it is unnecessary to add 
an extra penalty term to the loss function. In other words, 
in this case, $\beta$ in Eq.~(\ref{eq:loss_hardware_Sz}) is set to 0.

At this point, the expectation values of various operators and the 
fidelities of the states have significantly improved, as shown in 
Figs.~\ref{fig:1D_all_eigenstates}(e)-(h).  
When we use the symmetry-preserving quantum circuit in 
Fig.~\ref{fig:lattice_and_quantum_circuits}(f) with 5 layers,
an even more accurate results can be got.
Figs.~\ref{fig:1D_all_eigenstates}(i)-(p) present the energy values 
and fidelities obtained during this process. These 16 eigenstates 
in 5 separate runs are computed, with each run targeting eigenstates 
that share the same particle number. This is something the hardware 
efficient ansatz cannot achieve, as it can only compute the first $n$
eigenstates in one run, always starting from the lowest energy state.
In contrast, the particle-number-preserving circuit allows us to find 
eigenstates with specific particle numbers, sorted by energy from 
low to high within that particle number sector.

\begin{figure*}
\includegraphics[width=1.0\textwidth]{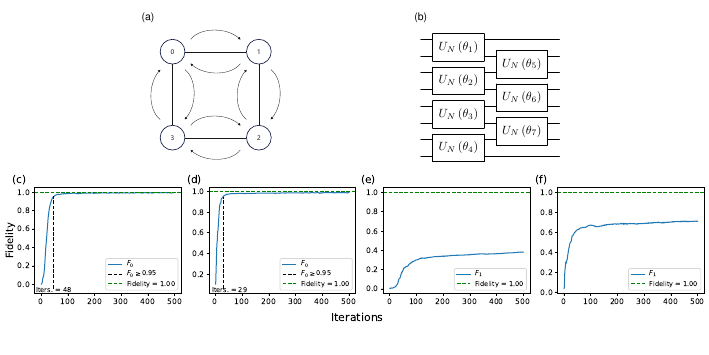}
\caption{\label{fig:lattice_and_circuit_4sites} (a) The one-dimensional Hubbard model with four lattice sites incorporates periodic boundary conditions, where circles represent lattice 
points, the numbers in the middle are the point indices, and arrows 
indicate that electrons only hop between nearest-neighbor lattice 
sites. (b) The particle number preserving circuit for the four-site
Hubbard model, applying 5 layers for the ground state calculation and 7
layers for the second excited state calculation. (c)-(d) The change of 
the average fidelity of hardware efficient ansatz and 
symmetry-preserving ansatz with the number of iterations when 
solving the ground state. The black vertical dashed line indicates 
the iteration number when the average fidelity first reaches 
$\geq0.95$. All data are the average of 5 independent computations. 
(e)-(f) Comparison between the two ansatzes when solving 
the second excited state.}
\end{figure*}

The loss function used when employing symmetry-preserving ansatz is
\begin{equation}
\mathcal{L}_{j}=\sum_{i=0}^{k_{j}-1}\omega_{ji}
\left[\langle\hat{H}\rangle_{ji}+\beta_{ji}\left(\langle\hat{S}_{z}\rangle_{ji}-M_{ji}\right)^{2}\right]\,,
\label{eq:loss_symmetry}
\end{equation}
where $j \in \{0, 1, ..., 4\}$ corresponds to different particle 
numbers. When $\langle\hat{S}_{z}\rangle_{ji}=0$, $\beta_{ji}=0$; 
and when $\langle\hat{S}_{z}\rangle_{ji}\neq0$, $\beta_{ji}=1$. 
The detailed expression is provided in Appendix 
~\ref{app:computational_resources}, where the iteration counts 
for achieving various fidelities by using three methods
are also compared. The hardware efficient ansatz without penalty 
terms requires 44–156 iterations to approach these eigenstates 
with a fidelity of 0.6, and some eigenstates fail to reach this 
fidelity within 300 iterations. Moreover, none of the eigenstates 
achieve an average fidelity of 0.9. In contrast, the 
particle-number-preserving ansatz with penalty terms requires only 
1–15 iterations to reach a fidelity of 0.6, 1–58 iterations to 
achieve a fidelity of 0.9, and 1–100 iterations to achieve 
a fidelity of 0.99 for most eigenstates. Some states converge to their 
corresponding energy eigenstates in the first iteration because their
initial states happen to coincide with those eigenstates. 
Note that the symmetry-preserving ansatz achieves these results with 
the same number of CNOT gates and significantly fewer parameters. 
The computational resources for different methods are provided in Appendix ~\ref{app:computational_resources}. 
This improvement is even more pronounced compared 
to the case of computing the ground state.

\section{\label{sec:partial_eigenstate}The partial eigenstate solutions of the four-site 1D Hubbard model}
In this section 
the one-dimensional Hubbard model with four lattice sites,
as shown in Fig.~\ref{fig:lattice_and_circuit_4sites}(a),
are considered. Extend the computational processes to a larger system.
This model incorporates periodic boundary conditions, which reduce the impact of boundary effects.
The arrows represent electron hopping between nearest-neighbor sites,
and the term $\sum_{i,j}\sum_{\sigma}\left(\hat{a}_{i\sigma}^{\dagger}\hat{a}_{j\sigma}+\hat{a}_{j\sigma}^{\dagger}\hat{a}_{i\sigma}\right)=\hat{H}_{0}$ in the Hamiltonian is expanded as follows
\begin{eqnarray}
\hat{H}_{0} & = & \left(\hat{a}_{0\uparrow}^{\dagger}\hat{a}_{1\uparrow}+\hat{a}_{1\uparrow}^{\dagger}\hat{a}_{0\uparrow}\right)+\left(\hat{a}_{0\uparrow}^{\dagger}\hat{a}_{3\uparrow}+\hat{a}_{3\uparrow}^{\dagger}\hat{a}_{0\uparrow}\right) \nonumber \\
& & +\left(\hat{a}_{1\uparrow}^{\dagger}\hat{a}_{2\uparrow}+\hat{a}_{2\uparrow}^{\dagger}\hat{a}_{1\uparrow}\right)+\left(\hat{a}_{2\uparrow}^{\dagger}\hat{a}_{3\uparrow}+\hat{a}_{3\uparrow}^{\dagger}\hat{a}_{2\uparrow}\right) \nonumber \\
& & +\left(\hat{a}_{0\downarrow}^{\dagger}\hat{a}_{1\downarrow}+\hat{a}_{1\downarrow}^{\dagger}\hat{a}_{0\downarrow}\right)+\left(\hat{a}_{0\downarrow}^{\dagger}\hat{a}_{3\downarrow}+\hat{a}_{3\downarrow}^{\dagger}\hat{a}_{0\downarrow}\right) \nonumber \\
& & +\left(\hat{a}_{1\downarrow}^{\dagger}\hat{a}_{2\downarrow}+\hat{a}_{2\downarrow}^{\dagger}\hat{a}_{1\downarrow}\right)+\left(\hat{a}_{2\downarrow}^{\dagger}\hat{a}_{3\downarrow}+\hat{a}_{3\downarrow}^{\dagger}\hat{a}_{2\downarrow}\right) .
\end{eqnarray}
The ground state and the second excited state of this system
are taken into acount. The first excited state was not included 
because it is a fourfold degenerate state, and these four degenerate
states cannot be distinguished solely based on particle number, 
total spin, or the $z$-component of spin. 
Nevertheless, since he other quantum numbers 
of the first excited states are unknown, it is not possible to match 
the degenerate states in computation with the theoretical 
eigenstates, thus making it challenging to solve effectively.

The ground state $\vert\iota_{2,0}\rangle$ has a particle number of 
2 and a $z$-component of spin equal to 0. The second excited state
$\vert\lambda_{4,0}\rangle$ has a particle number of 4 and a 
$z$-component of spin equal to 0. The initial states used to solve these 
eigenstates are equal-weight superpositions of all computational basis 
states satisfying these quantum numbers, with positive real 
coefficients. When solving for the second excited state, 
the hardware efficient ansatz requires solving for the lower 
eigenstates simultaneously, whereas the symmetry ansatz allows solving
for the second excited state directly. All data represent the averages
of 5 independent optimization processes.

A 15-layer hardware efficient ansatz are
employed (Fig.~\ref{fig:lattice_and_quantum_circuits}(b) 
extended to 8 qubits) to solve the ground state of this system, 
involving 240 parameters and 105 CNOT gates. 
For comparison, a 5-layer symmetry-preserving ansatz 
(Fig.~\ref{fig:lattice_and_circuit_4sites}(b)) was used, with 35 
parameters and 105 CNOT gates. The results show that the hardware 
efficient ansatz achieved a fidelity $\geq 0.95$ on average at the 
48th iteration, while the symmetry-preserving ansatz required only 29 
iterations, as illustrated in Fig.~\ref{fig:lattice_and_circuit_4sites}
(c)-(d). Notably, the symmetry-preserving ansatz achieved this result
using significantly fewer parameters.

When to solve the second excited state, the difference becomes even
more evident. A 21-layer hardware efficient ansatz 
without penalty terms (336 parameters and 147 CNOT gates) and a 7-layer 
symmetry-preserving ansatz with penalty terms (49 parameters and 147 
CNOT gates) were employed to solve this eigenstate. The former ansatz achieved an average fidelity of approximately 0.4 at the 500th 
iteration, whereas the latter ansatz reached a fidelity
of about 0.7, 
as shown in Fig.~\ref{fig:lattice_and_circuit_4sites}(e)-(f).
The achievable precision is limited by hardware constraints. Although the optimal fidelity was not reached, the comparison mainly serves to demonstrate the advantage of the symmetry ansatz, which is evident from the results.

\section{\label{sec:Conclusion} Conclusion}

Along with the increase of the  particle number of a system, the 
dimension of the Hilbert space of a quantum many-body system 
grows exponentially. For such systems, classical computers encounter
significant challenges in simulation when the system size becomes
sufficiently large. Quantum computers were proposed in the context of 
quantum simulation, leveraging the quantum nature of qubits to
naturally simulate other quantum systems. In the current NISQ era, 
the VQE is an effective algorithm for finding the energy eigenstates 
of quantum systems. Over time, VQE has evolved from solving only the
ground state to calculating excited states and from ignoring system
symmetries to efficiently incorporating them. 

Based on this foundation,
symmetry-enhanced VQE is employed in this paper to solve the energy eigenstates of the one-dimensional Fermi-Hubbard model with two and four sites, demonstrating the efficiency of symmetry methods in solving the Hubbard model. 
Three approaches are compared, 
which are: the hardware efficient ansatz that completely ignores 
symmetry, the hardware efficient ansatz that partially considers 
symmetry through the addition of penalty terms, and the 
symmetry-preserving ansatz. 
The third approach preserves the particle number of the state during 
the computation, and a penalty term was added for the $z$-component of
the spin operator in the loss function. Our results show that symmetry
methods can achieve better outcomes with fewer computational resources 
when solving the energy eigenstates of the Hubbard model. 
This suggests that for solving large-scale Hubbard models,
symmetries should be analyzed of the system to obtain effective results
with reduced analytical effort and numerical computational resources.
In future research, when computing higher-dimensional Hubbard models, we should also consider the symmetry of the system to verify the effectiveness of this method and apply it to more specific physical systems.

\section*{\label{app:Data_Availability} Data Availability}

All the derivation details, the code used for the simulation program, and the generated data are available at Ref. \cite{ZenodoOpenSource}.

\begin{acknowledgments}
This work was supported by the National Natural Science Foundation of China under Grant No. 12475104.

\end{acknowledgments}

\appendix

\section{\label{app:An_estimation} An estimation of the maximum matrix dimension that Frontier can solve}

The Frontier supercomputer has 74 racks, each with 64 blades, 
and each blade contains two computing nodes, for a total of 9,472 
nodes. Together, these nodes are equipped with 9,472 AMD EPYC
processors and 37,888 AMD Radeon Instinct Mi250X general purpose
graphics processing units (GPGPUs). Each processor and GPGPU has 
512GB and 128GB of memory, respectively, and each node has 5TB of 
non-volatile flash memory. Assuming each matrix element is a 64-bit
floating-point number, which occupies 8 bytes, it can be calculated
that the matrix size that Frontier can store is 
$8.84\times10^{7}\approx10^{8}$ dimensions.

\section{\label{app:SO3_SU2} SO(3) group element represented by SU(2) group parameters}

For the SU(2) group parameters $a,b\in\mathbb{C}$, with 
$\left|a\right|^{2}+\left|b\right|^{2}=1$, the matrix representation 
of the SO(3) group element in terms of these group parameters is 
given by
\begin{equation}
R\left(a,b\right)=\begin{pmatrix}\alpha & \beta & \gamma\\
\delta & \epsilon & \varepsilon\\
\zeta & \eta & \theta
\end{pmatrix}\,,
\end{equation}
where
\begin{eqnarray*}
\alpha & = & \frac{1}{2}\left(a^{2}+a^{*2}-b^{2}-b^{*2}\right) \,, \\
\beta & = & -\frac{\mathrm{i}}{2}\left(a^{2}-a^{*2}+b^{2}-b^{*2}\right) \,, \\
\gamma & = & -\left(ab+a^{*}b^{*}\right) \,, \\
\delta & = & \frac{\mathrm{i}}{2}\left(a^{2}-a^{*2}-b^{2}+b^{*2}\right) \,, \\
\epsilon & = & \frac{1}{2}\left(a^{2}+a^{*2}+b^{2}+b^{*2}\right) \,, \\
\varepsilon & = & \mathrm{i}\left(a^{*}b^{*}-ab\right) \,, \\
\zeta & = & a^{*}b+ab^{*} \,, \\
\eta & = & \mathrm{i}\left(a^{*}b-ab^{*}\right) \,, \\
\theta & = & aa^{*}-bb^{*} \,.
\end{eqnarray*}

\section{\label{app:Derivation_steps} Derivation steps for operators encoding}

The encodings of the fermion annihilation and creation operators for
spin-up and spin-down at the $i$-th lattice site are represented as
follows
\begin{eqnarray}
\hat{a}_{i\uparrow} & = & \prod_{k=0}^{i-1}Z_{k}\otimes\sigma_{i}^{-}\otimes\prod_{k=i+1}^{2N-1}I_{k} \,, \\
\hat{a}_{i\uparrow}^{\dagger} & = & \prod_{k=0}^{i-1}Z_{k}\otimes\sigma_{i}^{+}\otimes\prod_{k=i+1}^{2N-1}I_{k} \,, \\
\hat{a}_{i\downarrow} & = & \prod_{k=0}^{N-1}I_{k}\otimes\prod_{k=N}^{i-1+N}Z_{k}\otimes\sigma_{i+N}^{-}\otimes\prod_{k=i+1+N}^{2N-1}I_{k} \,, \\
\hat{a}_{i\downarrow}^{\dagger} & = & \prod_{k=0}^{N-1}I_{k}\otimes\prod_{k=N}^{i-1+N}Z_{k}\otimes\sigma_{i+N}^{+}\otimes\prod_{k=i+1+N}^{2N-1}I_{k} \,.
\end{eqnarray}

The following relations are used in the derivation:
\begin{eqnarray}
\sigma^{+}\cdot\sigma_{z} & = & \sigma^{+} \,, \\
\sigma_{z}\cdot\sigma^{-} & = & \sigma^{-} \,, \\
\sigma_{x}\cdot\sigma_{x} & = & I \,, \\
\sigma_{y}\cdot\sigma_{y} & = & I \,, \\
\sigma_{z}\cdot\sigma_{z} & = & I \,, \\
\sigma_{x}\cdot\sigma_{y} & = & \mathrm{i}\sigma_{z} \,, \\
\sigma_{y}\cdot\sigma_{x} & = & -\mathrm{i}\sigma_{z} \,,
\end{eqnarray}
as well as
\begin{eqnarray}
\sigma^{-} & = & \frac{1}{2}\left(\sigma_{x}+\mathrm{i}\sigma_{y}\right) \,, \\
\sigma^{+} & = & \frac{1}{2}\left(\sigma_{x}-\mathrm{i}\sigma_{y}\right) \,.
\end{eqnarray}
Given $ i < j $, we have
\begin{eqnarray}
\hat{a}_{i\uparrow}^{\dagger}\hat{a}_{j\uparrow} & = & \left(\prod_{k=0}^{i-1}Z_{k}\otimes\sigma_{i}^{+}\otimes\prod_{k=i+1}^{2N-1}I_{k}\right)\nonumber \\
 &  & \cdot\left(\prod_{k=0}^{j-1}Z_{k}\otimes\sigma_{j}^{-}\otimes\prod_{k=j+1}^{2N-1}I_{k}\right) \nonumber \\
 & = & \frac{1}{4}\prod_{k=0}^{i-1}I_{k}\otimes\left(X_{i}\otimes\prod_{k=i+1}^{j-1}Z_{k}\otimes X_{j}\right.\nonumber \\
 &  & +X_{i}\otimes\prod_{k=i+1}^{j-1}Z_{k}\otimes \mathrm{i}Y_{j}-\mathrm{i}Y_{i}\otimes\prod_{k=i+1}^{j-1}Z_{k}\otimes X_{j}\nonumber \\
 &  & \left.-\mathrm{i}Y_{i}\otimes\prod_{k=i+1}^{j-1}Z_{k}\otimes \mathrm{i}Y_{j}\right)\otimes\prod_{k=j+1}^{2N-1}I_{k} \,.
\end{eqnarray}
Note that for two vector spaces $ R_{1}, R_{2} $, and operators $ A, B \in R_{1} $, $ L, M \in R_{2} $, the relation $ \left(A \otimes L\right)\left(B \otimes M\right) = AB \otimes LM $ holds.

Additionally, since $ \hat{a}_{j\uparrow}^{\dagger}\hat{a}_{i\uparrow} = \left(\hat{a}_{i\uparrow}^{\dagger}\hat{a}_{j\uparrow}\right)^{\dagger} $, the encoding for $ \hat{a}_{i\uparrow}^{\dagger}\hat{a}_{j\uparrow} + \hat{a}_{j\uparrow}^{\dagger}\hat{a}_{i\uparrow} $ can be obtained. Similarly, the encoding for $ \hat{a}_{i\downarrow}^{\dagger}\hat{a}_{j\downarrow} + \hat{a}_{j\downarrow}^{\dagger}\hat{a}_{i\downarrow} $ can also be derived.

For terms like $ X_{i} \otimes \prod_{k=i+1}^{j-1}Z_{k} \otimes X_{j} $, consider the following explanation:  
First, $ i < j $. When $ j = i+1 $, we have 
$
X_{i} \otimes \prod_{k=i+1}^{j-1}Z_{k} \otimes X_{j} = X_{i} \otimes X_{i+1}.
$
When $ j = i+2 $, 
$
X_{i} \otimes \prod_{k=i+1}^{j-1}Z_{k} \otimes X_{j} = X_{i} \otimes Z_{i+1} \otimes X_{i+2},
$
and so on.

For terms like $ \prod_{k=0}^{i-1}I_{k} $:  
When $ i = 0 $, the term $ \prod_{k=0}^{i-1}I_{k} $ does not exist, leading to 
$
\hat{a}_{0\uparrow}^{\dagger}\hat{a}_{j\uparrow} + \hat{a}_{j\uparrow}^{\dagger}\hat{a}_{0\uparrow} = \frac{1}{2} \left(X_{0} \otimes \prod_{k=1}^{j-1}Z_{k} \otimes X_{j} + Y_{0} \otimes \prod_{k=1}^{j-1}Z_{k} \otimes Y_{j}\right) \otimes \prod_{k=j+1}^{2N-1}I_{k}.
$
When $ i = 1 $, $ \prod_{k=0}^{i-1}I_{k} = I_{0} $, and so forth. The term $ \prod_{k=j+1}^{2N-1}I_{k} $ follows a similar pattern.

For $ \hat{n}_{i\uparrow}\hat{n}_{i\downarrow} $,
\begin{eqnarray}
\hat{n}_{i\downarrow} & = & \hat{a}_{i\downarrow}^{\dagger}\hat{a}_{i\downarrow} \nonumber \\
 & = & \frac{1}{2}\prod_{k=0}^{i-1+N}I_{k}\otimes\left(I_{i+N}-Z_{i+N}\right) \nonumber \\
 &  & \otimes\prod_{k=i+1+N}^{2N-1}I_{k}\,.
\end{eqnarray}
Similarly, we can derive $ \hat{n}_{i\uparrow} = \hat{a}_{i\uparrow}^{\dagger}\hat{a}_{i\uparrow} $ and $ \hat{n}_{i\uparrow}\hat{n}_{i\downarrow} = \hat{a}_{i\uparrow}^{\dagger}\hat{a}_{i\uparrow}\hat{a}_{i\downarrow}^{\dagger}\hat{a}_{i\downarrow} $. 

In summary, we have
\begin{eqnarray}
\hat{a}_{i\uparrow}^{\dagger}\hat{a}_{j\uparrow}+\hat{a}_{j\uparrow}^{\dagger}\hat{a}_{i\uparrow} & = & \frac{1}{2}\prod_{k=0}^{i-1}I_{k}\otimes\left(X_{i}\otimes\prod_{k=i+1}^{j-1}Z_{k}\otimes X_{j}\right. \nonumber \\
 &  & \left.+Y_{i}\otimes\prod_{k=i+1}^{j-1}Z_{k}\otimes Y_{j}\right) \nonumber \\
 & & \otimes\prod_{k=j+1}^{2N-1}I_{k} \,,
\end{eqnarray}

\begin{eqnarray}
\hat{a}_{i\downarrow}^{\dagger}\hat{a}_{j\downarrow}+\hat{a}_{j\downarrow}^{\dagger}\hat{a}_{i\downarrow} & = & \frac{1}{2}\prod_{k=0}^{i-1+N}I_{k}\otimes\left(X_{i+N}\right. \nonumber \\
 &  & \otimes\prod_{k=i+1+N}^{j-1+N}Z_{k}\otimes X_{j+N} \nonumber \\
 &  & \left.+Y_{i+N}\otimes\prod_{k=i+1+N}^{j-1+N}Z_{k}\otimes Y_{j+N}\right) \nonumber \\
 &  & \otimes\prod_{k=j+1+N}^{2N-1}I_{k} \,,
\end{eqnarray}

\begin{eqnarray}
\hat{n}_{i\uparrow}\hat{n}_{i\downarrow} & = & \frac{1}{4}\prod_{k=0}^{i-1}I_{k}\otimes\left(I_{i}\otimes\prod_{k=i+1}^{i-1+N}I_{k}\otimes I_{i+N}\right. \nonumber \\
 &  & -I_{i}\otimes\prod_{k=i+1}^{i-1+N}I_{k}\otimes Z_{i+N} \nonumber \\
 &  & -Z_{i}\otimes\prod_{k=i+1}^{i-1+N}I_{k}\otimes I_{i+N} \nonumber \\
 &  & \left.+Z_{i}\otimes\prod_{k=i+1}^{i-1+N}I_{k}\otimes Z_{i+N}\right) \nonumber \\
 &  & \otimes\prod_{k=i+1+N}^{2N-1}I_{k} \,.
\end{eqnarray}

For $\hat{N} = \sum_{i,\sigma \in \{\uparrow, \downarrow\}} \hat{a}_{i\sigma}^\dagger \hat{a}_{i\sigma} = \sum_{i} \left(\hat{a}_{i\uparrow}^\dagger \hat{a}_{i\uparrow} + \hat{a}_{i\downarrow}^\dagger \hat{a}_{i\downarrow}\right)$ and $\hat{S}_{z} = \frac{1}{2}\sum_{i}\left(\hat{n}_{i\uparrow} - \hat{n}_{i\downarrow}\right)$, their encodings are given by
\begin{eqnarray}
\hat{N} & = & \frac{1}{2}\sum_{i}\prod_{k=0}^{i-1}I_{k}\otimes\left(I_{i}\otimes\prod_{k=i+1}^{i+N}I_{k}\right. \nonumber \\
 &  & -Z_{i}\otimes\prod_{k=i+1}^{i+N}I_{k}+\prod_{k=i}^{i-1+N}I_{k}\otimes I_{i+N} \nonumber \\
 &  & \left.-\prod_{k=i}^{i-1+N}I_{k}\otimes Z_{i+N}\right)\otimes\prod_{k=i+1+N}^{2N-1}I_{k} \,,
\end{eqnarray}
and
\begin{eqnarray}
\hat{S}_{z} & = & \frac{1}{4}\sum_{i}\prod_{k=0}^{i-1}I_{k}\otimes\left(I_{i}\otimes\prod_{k=i+1}^{i+N}I_{k}\right. \nonumber \\
 &  & -Z_{i}\otimes\prod_{k=i+1}^{i+N}I_{k}-\prod_{k=i}^{i-1+N}I_{k}\otimes I_{i+N} \nonumber \\
 &  & \left.+\prod_{k=i}^{i-1+N}I_{k}\otimes Z_{i+N}\right)\otimes\prod_{k=i+1+N}^{2N-1}I_{k} \,.
\end{eqnarray}

\section{\label{app:computational_resources} Computational resources, loss functions, initial states, and fidelities for the two-site Hubbard model}

\subsection{\label{app:sub_computational_resources} Computational resources}

The computational resources for the one-dimensional Hubbard model with two lattice sites using different methods are shown in the table below.

\begin{table}[H]
\caption{\label{tab:computational_resources_ground}
The number of parameters and CNOT gates used for solving the ground state with different ansatzes.
}
\begin{ruledtabular}
\begin{tabular}{lcc}
\textrm{Ansatz} & \textrm{Parameters} & \textrm{CNOT gates} \\
\colrule
Hardware efficient        & 96 & 36 \\
Symmetry-preserving type 1 & 48 & 36 \\
Symmetry-preserving type 2 & 12 & 36 \\
\end{tabular}
\end{ruledtabular}
\end{table}
\begin{table}[H]
\caption{\label{tab:computational_resources_excited}
The number of parameters and CNOT gates used for solving all energy eigenstates with different ansatzes. When using hardware efficient ansatz, the computational resources listed in the table remain consistent regardless of whether penalty terms are added to the loss function.
}
\begin{ruledtabular}
\begin{tabular}{lcc}
\textrm{Ansatz} & \textrm{Parameters} & \textrm{CNOT gates} \\
\colrule
Hardware efficient        & 120 & 45 \\
Symmetry-preserving       & 15  & 45 \\
\end{tabular}
\end{ruledtabular}
\end{table}

\subsection{\label{app:loss functions} Loss functions}
In Eq.~(\ref{eq:loss_hardware_Sz}), $\omega_{i}\in\left\{ 1,2,...,16\right\} $ and $\omega_{0}>\omega_{1}>\cdots>\omega_{k-1}$. When $i\in\left\{ 0,4,6,11,14,15\right\} $, $\beta_{i}=0$, and when $i\in\left\{ 1,2,3,5,7,8,9,10,12,13\right\} $, $\beta_{i}=0.5$. $\langle\hat{H}\rangle_{i}=\langle\psi_{i}\vert U^{\dagger}\left(\boldsymbol{\theta}\right)\hat{H}U\left(\boldsymbol{\theta}\right)\vert\psi_{i}\rangle$, $\langle\hat{S}_{z}\rangle_{i}=\langle\psi_{i}\vert U^{\dagger}\left(\boldsymbol{\theta}\right)\hat{S}_{z}U\left(\boldsymbol{\theta}\right)\vert\psi_{i}\rangle$. $M_{i}$ corresponds to Table~\ref{tab:eigenstates}, the $i$-th $z$-component of spin from top to bottom.

The specific expression for Eq.~(\ref{eq:loss_symmetry}) is
\begin{eqnarray}
\mathcal{L}_{j=0} & = & \langle\hat{H}\rangle_{4} \,, \\
\mathcal{L}_{j=1} & = & 4\left[\langle\hat{H}\rangle_{1}+\left(\langle\hat{S}_{z}\rangle_{1}+0.5\right)^{2}\right] \nonumber \\
& & +3\left[\langle\hat{H}\rangle_{2}+\left(\langle\hat{S}_{z}\rangle_{2}-0.5\right)^{2}\right] \nonumber \\
& & +2\left[\langle\hat{H}\rangle_{8}+\left(\langle\hat{S}_{z}\rangle_{8}-0.5\right)^{2}\right] \nonumber \\
& & +\left[\langle\hat{H}\rangle_{9}+\left(\langle\hat{S}_{z}\rangle_{9}+0.5\right)^{2}\right] \,, \\
\mathcal{L}_{j=2} & = & 6\langle\hat{H}\rangle_{0}+5\left[\langle\hat{H}\rangle_{3}+\left(\langle\hat{S}_{z}\rangle_{3}+1\right)^{2}\right] \nonumber \\
& & +4\left[\langle\hat{H}\rangle_{5}+\left(\langle\hat{S}_{z}\rangle_{5}-1\right)^{2}\right] \nonumber \\
& & +3\langle\hat{H}\rangle_{6}+2\langle\hat{H}\rangle_{11}+\langle\hat{H}\rangle_{14} \label{eq:Lj2} \,, \\
\mathcal{L}_{j=3} & = & 4\left[\langle\hat{H}\rangle_{7}+\left(\langle\hat{S}_{z}\rangle_{7}+0.5\right)^{2}\right] \nonumber \\
& & +3\left[\langle\hat{H}\rangle_{10}+\left(\langle\hat{S}_{z}\rangle_{10}-0.5\right)^{2}\right] \nonumber \\
& & +2\left[\langle\hat{H}\rangle_{12}+\left(\langle\hat{S}_{z}\rangle_{12}+0.5\right)^{2}\right] \nonumber \\
& & +\left[\langle\hat{H}\rangle_{13}+\left(\langle\hat{S}_{z}\rangle_{13}-0.5\right)^{2}\right] \,, \\
\mathcal{L}_{j=4} & = & \langle\hat{H}\rangle_{15} \,.
\end{eqnarray}

\begin{figure*}
\includegraphics[width=1.0\textwidth]{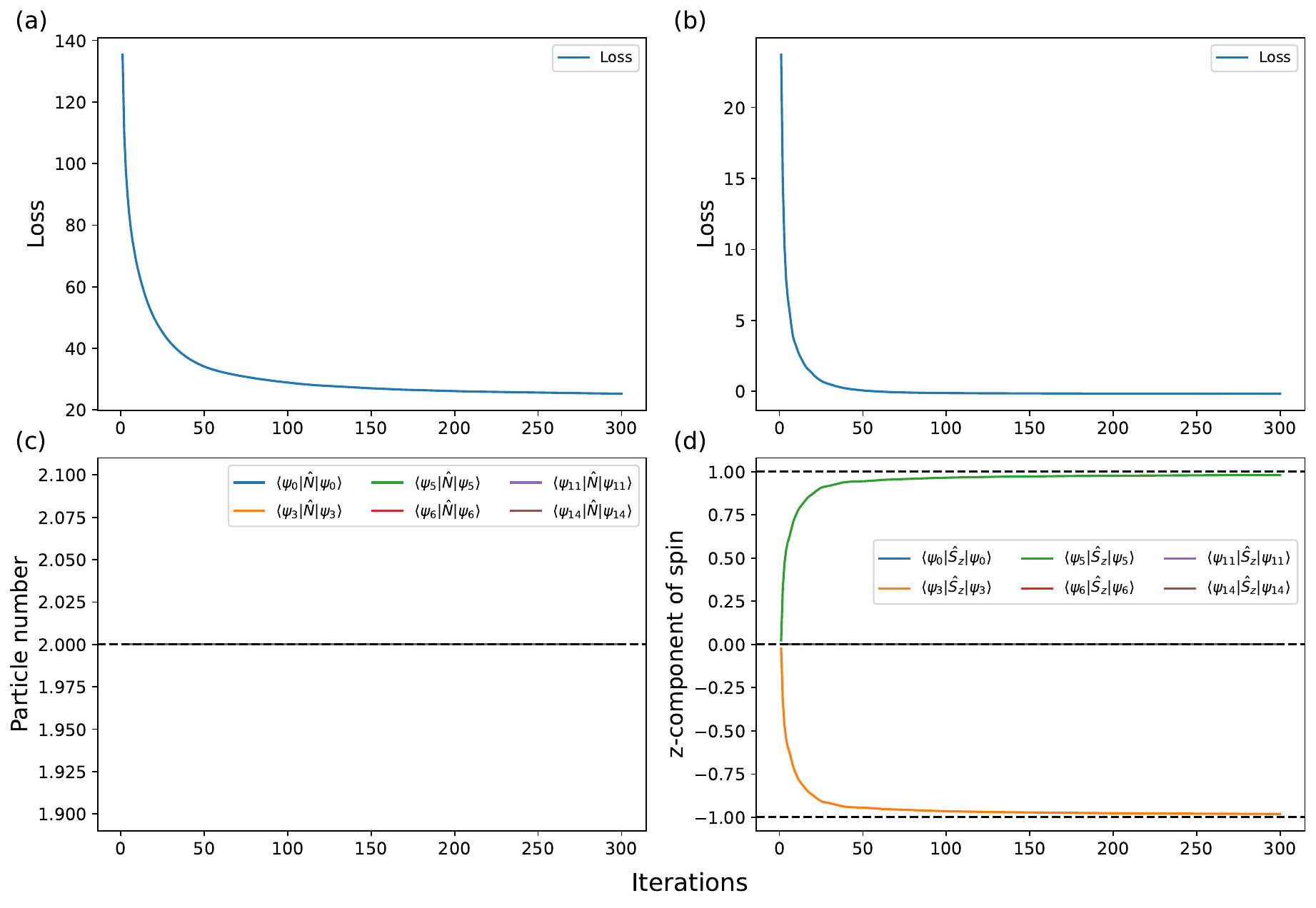}
\caption{\label{fig:1D_loss} Partial optimization processes of two 
different ansatz. (a) Variation of the loss function with the number 
of iterations for a hardware efficient ansatz without penalty terms. 
(b) Variation of the loss function with the number of iterations for a 
symmetry-preserving ansatz with penalty terms, where the initial
state's particle number $\langle\hat{N}\rangle$ is 2. (c) Particle
number of the state during the optimization process with the 
symmetry-preserving ansatz, showing that the circuit maintains a
particle number of 2. (d) $z$-component of spin
$\langle\hat{S}_{z}\rangle$ of the state during the optimization
process with the symmetry-preserving ansatz, demonstrating that the 
circuit maintains the $z$-component of spin when the initial state's 
$z$-component of spin is 0.}
\end{figure*}

Fig.~\ref{fig:1D_loss}(a) presents the loss function during the 
optimization process for a hardware efficient ansatz without
penalty terms (corresponding to the loss function in Eq.~(\ref{eq:loss_excited_state})). 
Fig.~\ref{fig:1D_loss}(b) shows the loss function during the 
computation of Eq.~(\ref{eq:Lj2}). From the figures, it can be observed 
that combining the symmetry-preserving circuit with penalty terms
allows the loss function to converge more quickly, becoming nearly 
constant after approximately 100 iterations. It is worth noting that
Fig.~\ref{fig:1D_loss}(a) considers all 16 eigenstates, while 
Fig.~\ref{fig:1D_loss}(b) includes only 6 eigenstates. 
This highlights the advantage of the symmetry-preserving circuit, 
as it is something the hardware efficient ansatz cannot achieve. 
The hardware efficient ansatz always starts from the lowest-energy 
eigenstate and works upward, requiring the simultaneous computation 
of the first $n$ eigenstates to obtain the $n$th eigenstate. In
contrast, the symmetry-preserving ansatz achieves rapid convergence
while calculating fewer eigenstates. Figs.~\ref{fig:1D_loss}(c)-(d)
depict the variation in particle number and $z$-component of spin during
the computation of Eq.~(\ref{eq:Lj2}), demonstrating that the circuit
maintains particle number invariance and, when the initial state's 
$z$-component of spin is 0, preserves the $z$-component of spin as well.

\subsection{\label{app:initial_states} Initial states}

The initial states used in the computation process are shown in Table~\ref{tab:eigenstates_initial_states}.

\begin{table}[H]
\caption{\label{tab:eigenstates_initial_states}
The 16 chosen mutually orthogonal initial states.
}
\begin{ruledtabular}
\begin{tabular}{l l}
\textrm{Eigenstate} & \textrm{Initial state} \\
\colrule
$\vert\alpha_{2,0}\rangle$ & $\vert\psi_{0}\rangle=\vert\psi^{+}\rangle\otimes\vert\psi^{+}\rangle$ \\
$\vert\beta_{1,-\frac{1}{2}}\rangle$ & $\vert\psi_{1}\rangle=\vert00\rangle\otimes\vert\psi^{+}\rangle$ \\
$\vert\beta_{1,\frac{1}{2}}\rangle$ & $\vert\psi_{2}\rangle=\vert\psi^{+}\rangle\otimes\vert00\rangle$ \\
$\vert\gamma_{2,-1}\rangle$ & $\vert\psi_{3}\rangle=\vert0011\rangle$ \\
$\vert\gamma_{0,0}\rangle$ & $\vert\psi_{4}\rangle=\vert0000\rangle$ \\
$\vert\gamma_{2,1}\rangle$ & $\vert\psi_{5}\rangle=\vert1100\rangle$ \\
$\vert\gamma_{2,0}\rangle$ & $\vert\psi_{6}\rangle=\vert\psi^{+}\rangle\otimes\vert\psi^{-}\rangle$ \\
$\vert\delta_{3,-\frac{1}{2}}\rangle$ & $\vert\psi_{7}\rangle=\vert\psi^{+}\rangle\otimes\vert11\rangle$ \\
$\vert\delta_{1,\frac{1}{2}}\rangle$ & $\vert\psi_{8}\rangle=\vert\psi^{-}\rangle\otimes\vert00\rangle$ \\
$\vert\delta_{1,-\frac{1}{2}}\rangle$ & $\vert\psi_{9}\rangle=\vert00\rangle\otimes\vert\psi^{-}\rangle$ \\
$\vert\delta_{3,\frac{1}{2}}\rangle$ & $\vert\psi_{10}\rangle=\vert11\rangle\otimes\vert\psi^{+}\rangle$ \\
$\vert\varepsilon_{2,0}\rangle$ & $\vert\psi_{11}\rangle=\vert\psi^{-}\rangle\otimes\vert\psi^{+}\rangle$ \\
$\vert\zeta_{3,-\frac{1}{2}}\rangle$ & $\vert\psi_{12}\rangle=\vert\psi^{-}\rangle\otimes\vert11\rangle$ \\
$\vert\zeta_{3,\frac{1}{2}}\rangle$ & $\vert\psi_{13}\rangle=\vert11\rangle\otimes\vert\psi^{-}\rangle$ \\
$\vert\eta_{2,0}\rangle$ & $\vert\psi_{14}\rangle=\vert\psi^{-}\rangle\otimes\vert\psi^{-}\rangle$ \\
$\vert\theta_{4,0}\rangle$ & $\vert\psi_{15}\rangle=\vert1111\rangle$ \\
\end{tabular}
\end{ruledtabular}
\end{table}

\subsection{\label{app:fidelities} The number of iterations corresponding to different fidelities}

The iteration numbers corresponding to different fidelities achieved
using various methods for the one-dimensional Hubbard model with two
lattice sites are shown in the tables below. 
Tables~\ref{tab:fidelity_0.6_hardware},~\ref{tab:fidelity_0.6_hardware_Sz}, and~\ref{tab:fidelity_0.6_symmetry} 
respectively present the iteration numbers when the average fidelity 
$\geq 0.6$ over 200 optimization steps, considering the hardware
efficient ansatz without penalty terms, the hardware efficient ansatz
with penalty terms, and the symmetry ansatz. The ``-'' indicates that,
during these 200 optimization steps, the state did not reach an average
fidelity of 0.6 within 300 iterations, as shown in 
Table~\ref{tab:fidelity_0.6_hardware}. 
Tables~\ref{tab:fidelity_0.9_hardware_Sz} 
and~\ref{tab:fidelity_0.9_symmetry} present the iteration numbers 
corresponding to an average fidelity $\geq 0.9$ for the hardware 
efficient ansatz with penalty terms and the symmetry ansatz, 
respectively. The hardware efficient ansatz without penalty terms 
fails to achieve an average fidelity of $\geq 0.9$ for any state 
within 300 iterations. To highlight the high performance of the
symmetry ansatz in solving eigenstates, we provide the iteration 
numbers for the symmetry ansatz when the average fidelity $\geq 0.99$, 
as shown in Table~\ref{tab:fidelity_0.99_symmetry}. Most eigenstates 
can be obtained within 100 iterations.

\vspace{-3mm}

\begin{table}[H]
\caption{\label{tab:fidelity_0.6_hardware}
Using the hardware efficient ansatz without the penalty term results 
in an average fidelity $\geq 0.6$.
}
\begin{ruledtabular}
\begin{tabular}{lcc}
\textrm{Fidelity type} & \textrm{Iteration} & \textrm{Fidelity threshold} \\
\colrule
$F_0$ & 68  & 0.6 \\
$F_1$ & 93  & 0.6 \\
$F_2$ & 95  & 0.6 \\
$F_3$ & -   & -   \\
$F_4$ & 108 & 0.6 \\
$F_5$ & 75  & 0.6 \\
$F_6$ & 156 & 0.6 \\
$F_7$ & -   & -   \\
$F_8$ & 90  & 0.6 \\
$F_9$ & -   & -   \\
$F_{10}$ & 99  & 0.6 \\
$F_{11}$ & 89  & 0.6 \\
$F_{12}$ & 112 & 0.6 \\
$F_{13}$ & 104 & 0.6 \\
$F_{14}$ & 91  & 0.6 \\
$F_{15}$ & 44  & 0.6 \\
\end{tabular}
\end{ruledtabular}
\end{table}

\vspace{-8mm}

\begin{table}[H]
\caption{\label{tab:fidelity_0.6_hardware_Sz}
Using a hardware efficient ansatz with a penalty term results 
in an average fidelity $\geq 0.6$.
}
\begin{ruledtabular}
\begin{tabular}{lcc}
\textrm{Fidelity type} & \textrm{Iteration} & 
\textrm{Fidelity threshold} \\
\colrule
$F_0$ & 26  & 0.6 \\
$F_1$ & 26  & 0.6 \\
$F_2$ & 26  & 0.6 \\
$F_3$ & 23  & 0.6 \\
$F_4$ & 29  & 0.6 \\
$F_5$ & 22  & 0.6 \\
$F_6$ & 50  & 0.6 \\
$F_7$ & 31  & 0.6 \\
$F_8$ & 30  & 0.6 \\
$F_9$ & 30  & 0.6 \\
$F_{10}$ & 32  & 0.6 \\
$F_{11}$ & 51  & 0.6 \\
$F_{12}$ & 32  & 0.6 \\
$F_{13}$ & 32  & 0.6 \\
$F_{14}$ & 34  & 0.6 \\
$F_{15}$ & 27  & 0.6 \\
\end{tabular}
\end{ruledtabular}
\end{table}


\begin{table}[H]
\caption{\label{tab:fidelity_0.6_symmetry}
Using a particle-number-preserving ansatz with a penalty term 
results in an average fidelity $\geq 0.6$.
}
\begin{ruledtabular}
\begin{tabular}{lcc}
\textrm{Fidelity type} & \textrm{Iteration} & \textrm{Fidelity threshold} \\
\colrule
$F_0$ & 4  & 0.6 \\
$F_1$ & 4  & 0.6 \\
$F_2$ & 6  & 0.6 \\
$F_3$ & 5  & 0.6 \\
$F_4$ & 1  & 0.6 \\
$F_5$ & 5  & 0.6 \\
$F_6$ & 15 & 0.6 \\
$F_7$ & 4  & 0.6 \\
$F_8$ & 7  & 0.6 \\
$F_9$ & 5  & 0.6 \\
$F_{10}$ & 5  & 0.6 \\
$F_{11}$ & 12 & 0.6 \\
$F_{12}$ & 6  & 0.6 \\
$F_{13}$ & 5  & 0.6 \\
$F_{14}$ & 3  & 0.6 \\
$F_{15}$ & 1  & 0.6 \\
\end{tabular}
\end{ruledtabular}
\end{table}


\begin{table}[H]
\caption{\label{tab:fidelity_0.9_hardware_Sz}
Using a hardware efficient ansatz with a penalty term results 
in an average fidelity $\geq 0.9$.
}
\begin{ruledtabular}
\begin{tabular}{lcc}
\textrm{Fidelity type} & \textrm{Iteration} & \textrm{Fidelity threshold} \\
\colrule
$F_0$ & 64  & 0.9 \\
$F_1$ & 65  & 0.9 \\
$F_2$ & 83  & 0.9 \\
$F_3$ & 56  & 0.9 \\
$F_4$ & 87  & 0.9 \\
$F_5$ & 57  & 0.9 \\
$F_6$ & 254 & 0.9 \\
$F_7$ & 127 & 0.9 \\
$F_8$ & 108 & 0.9 \\
$F_9$ & 81  & 0.9 \\
$F_{10}$ & 95  & 0.9 \\
$F_{11}$ & 244 & 0.9 \\
$F_{12}$ & 100 & 0.9 \\
$F_{13}$ & 88  & 0.9 \\
$F_{14}$ & 90  & 0.9 \\
$F_{15}$ & 64  & 0.9 \\
\end{tabular}
\end{ruledtabular}
\end{table}


\begin{table}
\caption{\label{tab:fidelity_0.9_symmetry}
Using a particle-number-preserving ansatz with a penalty term 
results in an average fidelity $\geq 0.9$.
}
\begin{ruledtabular}
\begin{tabular}{lcc}
\textrm{Fidelity type} & \textrm{Iteration} & 
\textrm{Fidelity threshold} \\
\colrule
$F_0$ & 19  & 0.9 \\
$F_1$ & 13  & 0.9 \\
$F_2$ & 18  & 0.9 \\
$F_3$ & 23  & 0.9 \\
$F_4$ & 1   & 0.9 \\
$F_5$ & 23  & 0.9 \\
$F_6$ & 58  & 0.9 \\
$F_7$ & 13  & 0.9 \\
$F_8$ & 19  & 0.9 \\
$F_9$ & 17  & 0.9 \\
$F_{10}$ & 15  & 0.9 \\
$F_{11}$ & 41  & 0.9 \\
$F_{12}$ & 19  & 0.9 \\
$F_{13}$ & 18  & 0.9 \\
$F_{14}$ & 8   & 0.9 \\
$F_{15}$ & 1   & 0.9 \\
\end{tabular}
\end{ruledtabular}
\end{table}


\begin{table}[H]
\caption{\label{tab:fidelity_0.99_symmetry}
Using a particle-number-preserving ansatz with a penalty term 
results in an average fidelity $\geq 0.99$.
}
\begin{ruledtabular}
\begin{tabular}{lcc}
\textrm{Fidelity type} & \textrm{Iteration} & 
\textrm{Fidelity threshold} \\
\colrule
$F_0$ & 70  & 0.99 \\
$F_1$ & 43  & 0.99 \\
$F_2$ & 45  & 0.99 \\
$F_3$ & -   & -    \\
$F_4$ & 1   & 0.99 \\
$F_5$ & -   & -    \\
$F_6$ & -   & -    \\
$F_7$ & 42  & 0.99 \\
$F_8$ & 53  & 0.99 \\
$F_9$ & 53  & 0.99 \\
$F_{10}$ & 44  & 0.99 \\
$F_{11}$ & 100 & 0.99 \\
$F_{12}$ & 51  & 0.99 \\
$F_{13}$ & 51  & 0.99 \\
$F_{14}$ & 31  & 0.99 \\
$F_{15}$ & 1   & 0.99 \\
\end{tabular}
\end{ruledtabular}
\end{table}

\nocite{}

\bibliography{Eigenstates_symmetry}  

\end{document}